\documentclass[letterpape,11pt]{article}
\pdfoutput = 1 
    
\usepackage{physics}    
\usepackage{jheppub} 
\usepackage[utf8]{inputenc}
\usepackage{amsmath}
\usepackage{slashed}
\usepackage{appendix}
\usepackage{caption}
\usepackage{subcaption}

\newcommand{\nl}{\nonumber \\[5pt]}

\newcommand{\bmat}[1]{\boldsymbol{#1}}

\newcommand{\lb}{\Big{\lbrack}}
\newcommand{\rb}{\Big{\rbrack}}
\newcommand{\lp}{\Big{(}}
\newcommand{\rp}{\Big{)}}
\newcommand{\lbc}{\Big{\lbrace}}
\newcommand{\rbc}{\Big{\rbrace}}
\newcommand{\Bvert}{\Big{\vert}}

\newcommand{\gcusp}{\gamma_{\text{cusp}} }

\newcommand{\eq}[1]{eq.~(\ref{eq:#1})}

\makeatletter
\def\mathcolor#1#{\@mathcolor{#1}}
\def\@mathcolor#1#2#3{%
  \protect\leavevmode
  \begingroup
    \color#1{#2}#3%
  \endgroup
}
\makeatother

\title{Transverse momentum dependent distributions in dijet and heavy hadron pair production at EIC}

\author[a]{Rafael F. del Castillo,}
\author[b,c]{Miguel G. Echevarria,}
\author[d]{Yiannis Makris}
\author[a]{and Ignazio Scimemi}
\affiliation[a]{Dpto. de F\'{i}sica Te\'{o}rica \& IPARCOS, Universidad Complutense de Madrid, E-28040 Madrid, Spain}
\affiliation[b]{
Department of Physics, University of the Basque Country UPV/EHU, Apartado 644, 48080 Bilbao, Spain}
\affiliation[c]{University of Alcalá, Dep. of Physics and Mathematics, 28805 Alcal\'{a} de Henares (Madrid), Spain}
\affiliation[d]{INFN Sezione di Pavia, via Bassi 6, I-27100 Pavia, Italy}

\emailAdd{raffer06@ucm.es}
\emailAdd{miguel.garciae@ehu.eus}
\emailAdd{yiannis.makris@pv.infn.it}
\emailAdd{ignazios@ucm.es}

\abstract{
We discuss the measurement of gluon transverse momentum distribution (TMD) in dijet and heavy hadron pair (HHP) production in semi-inclusive deep inelastic scattering. The factorization of these processes in position space shows the appearance of a specific new soft factor matrix element on top of angular and complex valued anomalous dimensions. We show in detail how these features can be treated consistently and we discuss a scale prescription  for the evolution kernel of the dijet soft function. As a result we obtain phenomenological predictions for unpolarized and angular modulated cross-sections for the electron-ion collider (EIC) using current available information on unpolarized TMD. 
}

\date{\today}

\begin{document}

\maketitle
\section{Introduction}
The access to non-perturbative gluon distributions from experiments is notoriously challenging. This is also the case of gluon transverse momentum distributions (TMDs). Gluons enter directly in Higgs production in hadronic colliders~\cite{Gao:2005iu,Chiu:2012ir, Echevarria:2015uaa,Neill:2015roa,Gutierrez-Reyes:2019rug} that has a relatively high mass and low production rates, and quarkonium production both at EIC and LHC
~\cite{Echevarria:2015uaa,Mulders:2000sh,Boer:2012bt,Ma:2012hh,Zhang:2014vmh,Ma:2015vpt, Boer:2015uqa,Bain:2016rrv,Mukherjee:2015smo,Mukherjee:2016cjw,Lansberg:2017tlc,Lansberg:2017dzg,Bacchetta:2018ivt,Hadjidakis:2018ifr,DAlesio:2019qpk,Echevarria:2019ynx,Fleming:2019pzj,Scarpa:2019fol,Grewal:2020hoc,Boer:2020bbd,Echevarria:2020qjk} that is sensitive also to the heavy quark hadronization effects \cite{Echevarria:2019ynx,Fleming:2019pzj}. 
Recent studies (see for example \cite{Page:2019gbf,AbdulKhalek:2021gbh}) suggest that the experimental observation of the dijet imbalance is possible at the future EIC.
In a recent work \cite{delCastillo:2020omr} we have proposed the dijet and hadron pair production at electron-ion colliders (EIC) to probe gluon TMD.  Previous studies on these processes were performed in ref.~\cite{Dominguez:2010xd,Hatta:2019ixj} for dijet and in ref.~\cite{Zhu:2013yxa, Zhang:2017uiz, Boer:2010zf} for heavy-meson pair production in an electron-hadron collider.
\begin{figure}
\begin{center}
\includegraphics[width=\textwidth]{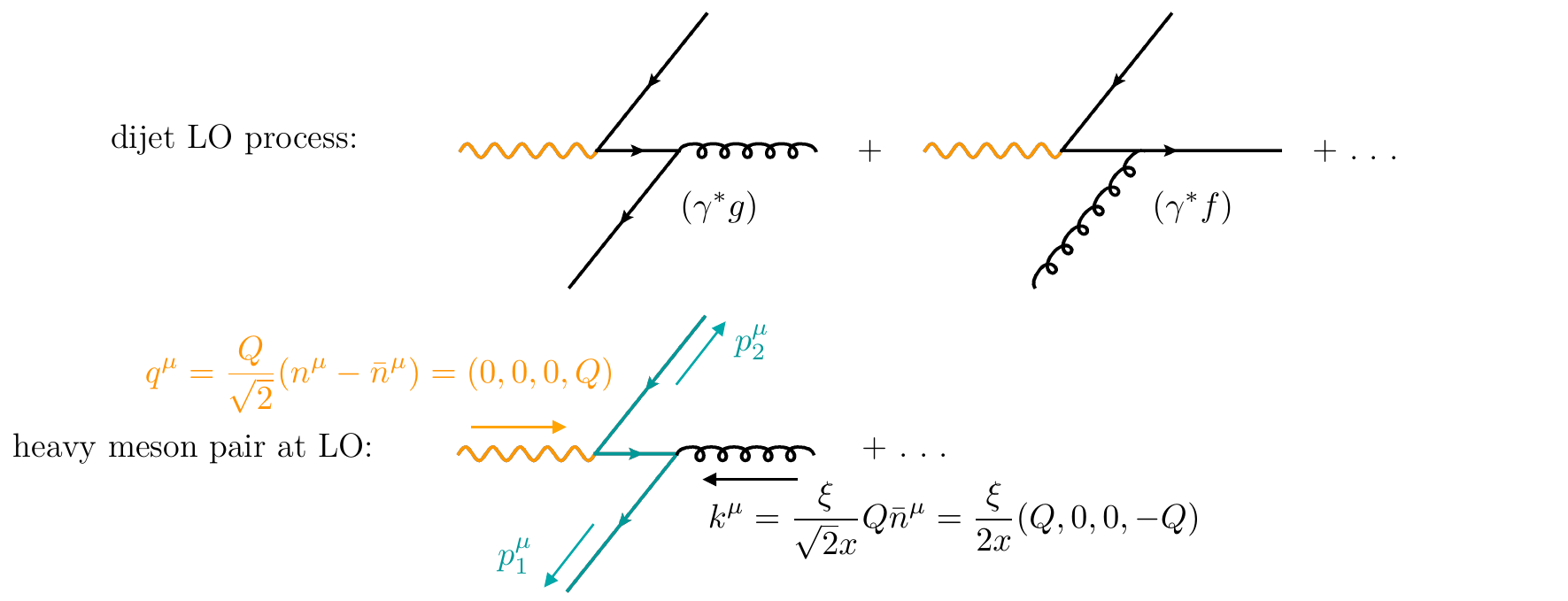}
\caption{\label{fig:LO} Example LO diagrams for the two processes. The momenta $q^{\mu}$ and $k^{\mu}$ (corresponding to the photon and incoming parton momenta respectively) are expressed in the Breit frame.}
\end{center}
\end{figure}
The relevant processes for our case are
\begin{align}\label{eq:1}
\ell + h&\to  \ell'+J_1 +J_2+X,\;&\text{and}& \; & \ell + h&\to  \ell'+ H + \bar{H}+X
\,,
\end{align}
where $\ell$ and $\ell'$ are the initial and final state leptons, $h$ is the colliding hadron, $J_i$ and $H/\bar{H}$ are the jets and heavy mesons respectively and $X$ represent undetected particles. At leading order (LO), and ignoring the intrinsic momentum of partons inside the target hadron, the two hard-scattering processes are schematically shown in fig.~\ref{fig:LO}. We have shown that these processes are factorizable when considering the cross-section
\begin{align}  
\frac{d\sigma
}{dx d\eta_1 d\eta_2 dp_T d\bmat{r}_T} 
\,,
\label{eq:2} 
\end{align}
where $x$ is the Bjorken variable, and $\eta_i$, $ \bmat{r}_T$  and $p_T$ are  respectively the rapidity, the sum of the transverse momenta (with respect to the beam axis) and the average scalar transverse momenta of the two final jets. The factorization condition is  $| \bmat{r}_T| \ll p_T$ in the Breit frame, as the virtual photon and target-hadron directions are back-to-back. These conditions are certainly fulfilled in the Breit frame for  $p_T\in [5,\, 40]$~GeV and  in the central rapidity region. We also demand that  there are no hierarchies among partonic Mandelstam variables, $\hat{s} \sim |\hat{t}| \sim |\hat{u}|$.
On the experimental side recent studies \cite{Page:2019gbf} suggest that the measurement of dijet imbalance is possible at EIC.
For the heavy-meson case monte-carlo generator studies  suggest that  charmed mesons can be reconstructed ~\cite{Arratia:2020azl, Chudakov:2016ytj}. The charm production rates have been considered at the LO and NLO QCD for $ep \to c/\bar{c} + X $ in ref.~\cite{Li:2020zbk}. For our case in principle we require the transverse momenta of the heavy mesons, $p_T^{H/\bar{H}}$, be parametrically larger than their mass, $m_H$, i.e. $p^{H/\bar{H}}_T \gg m_H$.

At leading order (LO) the jet processes can be initiated by 
either a gluon or a quark, while in the heavy-meson case only the gluon initial state is relevant. The factorized cross-section results in products/convolutions of several fundamental functions as TMDs, jet/heavy quark distributions, soft function and a new evolution kernel, detailed in \cite{delCastillo:2020omr}. The purpose of the present paper is to provide a phenomenological study of these processes, including theoretical errors. In order to achieve this, we have used the code Artemide \cite{web},
introducing new moduli necessary to describe the present cases. The code already includes quark TMDPDF  and the TMD evolution kernel extracted from Drell-Yan and semi-inclusive DIS experiments \cite{Scimemi:2019cmh}. 

The factorization of the cross-section is produced in position space, described by the variable $\bmat{b}$, conjugate of the momentum $\bmat{r}_T$. In order to Fourier transform the factorized cross-section from $\bmat{b}$ space to momentum space one has to perform an angular integration on $\phi_b$  (i.e. $d^2\bmat{b}=b\, db\,d\phi_b$ and  $\bmat v_J \cdot \bmat b = v_J b \cos \phi_b $) that results non trivial because the anomalous dimensions of several functions depend on this angle and are complex valued. We show that this integration can be performed in resummed perturbation theory and that this addresses the problem of complex values in all anomalous dimensions.
As a result the  $\phi_b$-angle integrated cross-section is then factorized into distributions that are 
 $\bmat{b}$-rotation invariant. 
The $\phi_b$-angle integrated dijet evolution kernel  is derived integrating a system of coupled differential equations similar to the TMD case. 
We show and discuss here a specific scale choice prescription that is analogue to the $\zeta$-prescription already discussed in \cite{Scimemi:2018xaf}, which was already implemented in Artemide.
 
The factorization theorem and the detailed definition of the observables is provided in sec.~\ref{sec:2}.
The cross-sections subtleties discovered when comparing to other groups   are discussed in sec.~\ref{sec:3}.
The resummation of logs involves angular integrations as  explained in sec.~\ref{sec:eka}, which leads to the evolution kernels detailed in 
sec.~\ref{sec:Evolution}. The phenomenological results obtained with the codes that we have developed are summarized in sec.~\ref{sec:results}, after which the conclusions are drawn.
\section{Factorization theorem, frame choice and modulations}
\label{sec:2}
\subsection{Notation and kinematics}
In order to define angles and to deduce a factorized cross-section we need to establish some kinematics.
The direction of the beam is fixed along  the $\hat{z}$ axis.  It is useful to define the four-vectors 
$n^{\mu} = \frac{1}{\sqrt{2}}(1,0,0,1),\;\bar n^{\mu} = \frac{1}{\sqrt{2}}(1,0,0,-1)
$,
so that $ n^2 = \bar{n}^2 = 0,\;  \bar{n}\cdot n = 1$.
Then any other four vector can be decomposed  into its light-cone components,  
 \begin{align}
    p^\mu &= p_+ \bar{n}^{\mu} + p_- n^{\mu} + p_{\perp}^{\mu} = (p_+,p_-,p_\perp)_{n},
\end{align}
with
\begin{align}
 p_+ = n\cdot p,\;\quad  p_- = \bar{n}\cdot p,\;\quad    p^2 = 2p_+ p_- + p^2_\perp = 2p_+ p_- - \bmat{p}^2.
\end{align}
The direction of the two jets are $v_1$ and $v_2$, normalized as 
\begin{align}
v_J^2 = \bar{ v}_J^2 = 0, \;\quad v_J\cdot \bar{v}_J  = 1,\;\text{ with }  J =1,2
\,, 
\end{align}
and  $\bar{v}_J$ are defined by reversing the sign of the spacial components. 
We have then the standard Lorentz-invariants,
\begin{align}
    Q^2 = -q^2,\; \quad x = \frac{Q^2}{2 P\cdot q},
\end{align}
where $q^{\mu}$ is the momentum of the virtual photon, $P^{\mu}$ is the momentum of the target hadron.  
In the Breit frame we have $q^{\mu}=(0,0,0,Q)$ and neglecting  mass corrections 
$P^{\mu} = \frac{1}{2x} (Q,0,0,-Q)\;.
$ 
The ratio of the longitudinal momenta of the incoming parton and the target hadron is
$
  \xi = \frac{k^+}{P^+}\;.
$
with  $k^{\mu}$ the momentum of the parton entering hard process. We can then express the variables $Q$ and $\xi$ in terms of the  Born level kinematics using the pseudo-rapidities, $\eta_1$ and $\eta_2$, and the transverse momentum, $p_T$, of the two outgoing partons,
\begin{align}
    Q &= 2p_T \cosh(\eta_-) \exp(\eta_+),\;& \xi = 2 x \cosh(\eta_+) \exp(-\eta_+)\;,
\end{align}
where, neglecting corrections from the target hadron mass,
$
    \eta_{\pm} = \frac{\eta_1 \pm \eta_2}{2}\;.
$ 
The partonic Mandelstam variables can be written using the same variables, 
\begin{align}
    \hat{s} &=(q+k)^2 =   4 p_T^2 \cosh^2(\eta_-) \;, \nl
    \hat{t} &=(q-p_2)^2= -4 p_T^2 \cosh(\eta_-) \cosh(\eta_+) \exp(\eta_1)\;, \nl
    \hat{u} &=(q-p_1)^2= -4 p_T^2 \cosh(\eta_-) \cosh(\eta_+) \exp(\eta_2)\;,
\end{align}
with $p_1^{\mu}$ and $p_2^{\mu}$  the momenta of the outgoing partons. At partonic level they satisfy
\begin{equation}
    \hat{s} +\hat{t}+\hat{u} = - Q^2\;.
\end{equation}

Finally, the transverse momentum imbalance of the two jets, $\bmat{r}_T$, and the hard transverse momentum, $p_T$, are defined through
\begin{align}
    \bmat{r}_T & = \bmat{p}_{1T} + \bmat{p}_{2T} ,\; &\bmat{p}_T  &= \frac{\bmat{p}_{1T}  - \bmat{p}_{2T} }{2},
\end{align}
where the sub-index 1, 2 refers to the final jets. 
At Born level $\bmat{p}_{1T} =  - \bmat{p}_{2T}$ and thus $ \bmat{r}_T  =0 $.  
It must be taken into account that the hadronization of the outgoing partons will form jet-like configurations along similar directions and wide angle radiation that can escape the jet clustering algorithm,  affecting the imbalance.
\subsection{Factorization theorem for dijet  and heavy hadron pair production}
The factorization of dijet  and heavy hadron pair production  at leading power (LP) for semi-inclusive deep inelastic experiments has already been provided in \cite{delCastillo:2020omr}.  The cross-sections reported here do not take into account any leptonic fiducial cuts, which however could be implemented once the experimental conditions are established (especially at EIC).
In this section we recall the main formulas that are used in our phenomenological description. We start with the dijet cross-section which can be written as  a sum of terms depending on the parton that initiates the hard process (quark or gluon) 
\begin{align}
 d\sigma_{2J}=    d & \sigma (\gamma^* g ) + d\sigma^U (\gamma^* f ), \\
 \label{eq:dsg}
  d & \sigma (\gamma^* g )= d\sigma^U (\gamma^* g ) + d\sigma^L (\gamma^* g ).
\end{align}
For an unpolarized hadronic process, the cross-section is made out of hard contributions from unpolarized  initial quarks $d\sigma^U (\gamma^* f )$,
unpolarized  initial gluons $d\sigma^U (\gamma^* g )$, and linearly polarized gluons  $d\sigma^L (\gamma^* f )$.
The quark contribution to the cross-section is
\begin{multline}\label{eq:FactGamFU}
    \frac{d\sigma^{U} (\gamma^* f)}{dx d\eta_1 d\eta_2 dp_T d\bmat{r}_T} = \sum_{f}\sigma_0^{fU}\, H^{U}_{\gamma^*f\to g f }(\hat{s},\hat{t}, \hat{u},\mu) \int \frac{d^2\bmat{b}}{(2\pi)^2}  \, \exp(i \bmat{b} \cdot \bmat{r}_T) \,f_{1}^{f}(\xi, \bmat{b},  \mu,\zeta_1) \\
    \times S_{\gamma f}(\bmat{b},\zeta_2,\mu) \lp \mathcal{C}_{g}(\bmat{b},R,\mu)  J_{g}(p_T,R,\mu) \rp  \lp \mathcal{C}_{f}(\bmat{b},R,\mu)  J_{f}(p_T,R,\mu) \rp.
\end{multline}
In this formula $f_1^{f}$ is the unpolarized quark TMDPDF for flavor $f$,
$H^U$ the hard factor for the unpolarized quark case.  The perturbative calculations of TMDPDF has been performed recently at NNLO \cite{Gehrmann:2014yya,Echevarria:2016scs,Gutierrez-Reyes:2019rug,Luo:2019hmp,Luo:2019bmw} and N3LO \cite{Luo:2019szz,Luo:2020epw}.
The jets are described by the product of a collinear-soft function $ \mathcal{C}_{(f,g)}$ and a jet shape function $J_{(f,g)}$ specific for each partonic flavor. 
The calculation at NLO of these functions can be found   in \cite{Hornig:2016ahz,Buffing:2018ggv} for generic $k_T$-type and cone jet algorithms.

The factor $S_{\gamma f}$ is the dijet soft function for the fundamental representation of SU(3)$_C$ and calculated in \cite{delCastillo:2020omr} up to NLO. A corresponding soft factor, $S_{\gamma g}$, for the adjoint representation of SU(3)$_C$ is also necessary for the incoming gluon contribution.
\begin{align}\label{eq:FactFormula}
    \frac{d\sigma (\gamma^* g)}{dx d\eta_1 d\eta_2 d p_T d\bmat{r}_T} &=
    \sum_{f} H^{ \mu \nu}_{\gamma^*g\to f \bar{f} }(\hat{s},\hat{t}, \hat{u},\mu) \int \frac{d^2 \bmat{b}}{(2\pi)^2}  \, \exp(i \bmat{b} \cdot \bmat{r}_T) \,F_{g, \mu \nu}(\xi, \bmat{b}, \mu,\zeta_1) \\ & \nonumber
    \times S_{\gamma g}(\bmat{b},\eta_1, \eta_2,\mu,\zeta_2) \,\lp \mathcal{C}_{f}(\bmat{b},R,\mu)  J_{f}(p_T,R,\mu) \rp \lp \mathcal{C}_{\bar{f}}(\bmat{b},R,\mu)  J_{\bar{f}}(p_T,R,\mu) \rp \;,
\end{align}
The hard factor $H^{ \mu \nu}(\mu)$ accounts for contributions of unpolarized and linearly polarized gluons,
\begin{equation}\label{eq:hadgtens}
    H^{\mu\nu}_{\gamma^* g \to f\bar{f}} = \sigma_0^{g U}\, H^U_{\gamma^*g\to f \bar{f} } \frac{g_{T}^{\mu\nu}}{d-2} + \sigma_0^{g L} \,H^L_{\gamma^*g\to f \bar{f} } \lp -\frac{g_{T}^{\mu\nu}}{d-2} +\frac{v^{\mu}_{1T}\, v^{\nu}_{2T} + v^{\mu}_{2T}\, v^{\nu}_{1T} }{2 \;v_{1T} \cdot v_{2T}}\rp 
    \,.
\end{equation}
The TMD tensor $F_{g, \mu \nu}$ can be also decomposed in terms of unpolarized and linearly polarized parts,
\begin{equation}
\label{eq:Fgtens}
    F_{g}^{\mu\nu}(\xi,\bmat{b}) = f^g_1(\xi,\bmat{b})\frac{g_{T}^{\mu\nu}}{d-2}  + h_1^\perp (\xi,\bmat{b})\,\lp \frac{g_{T}^{\mu\nu}}{d-2} + \frac{b^{\mu} b^{\nu}}{\bmat{b}^2}\rp
\,,
\end{equation}
with $g_T^{\mu\nu}=g^{\mu\nu}-n^\mu\bar n^\nu-\bar n^\mu n^\nu$. The hard factors are evaluated up to NNLO in the unpolarized case in \cite{Becher:2009th,Becher:2012xr} and at LO  for the linearly polarized case \cite{Chien:2020hzh}. 
In these equations $f^g_1$ and $h_1^\perp$ represent the unpolarized and linearly polarized gluon TMD. Both of them are known perturbatively up to NNLO~\cite{Echevarria:2016scs,Gutierrez-Reyes:2019rug,Luo:2019bmw}.  Combining eq.~(\ref{eq:dsg}, \ref{eq:FactFormula}, \ref{eq:hadgtens}, \ref{eq:Fgtens}) one obtains
\begin{multline}\label{eq:FactGamGU}
    \frac{d\sigma^U  (\gamma^* g )}{dx d\eta_1 d\eta_2 dp_T d\bmat{r}_T} = \sigma_0^{g U}\,\sum_{f} H^{U}_{\gamma^*g\to f \bar{f} }(\hat{s},\hat{t}, \hat{u},\mu) \int \frac{d^2\bmat{b}}{(2\pi)^2}  \, \exp(i \bmat{b} \cdot \bmat{r}_T) \, f_1^g(\xi, \bmat{b}, \mu,\zeta_1) \\
    \times S_{\gamma g}(\bmat{b},\zeta_2,\mu) \,\lp \mathcal{C}_{f}(\bmat{b},R,\mu)  J_{f}(p_T,R,\mu) \rp \lp \mathcal{C}_{\bar{f}}(\bmat{b},R,\mu)  J_{\bar{f}}(p_T,R,\mu) \rp\;,    
\end{multline}
\begin{multline}\label{eq:FactGamGL}
    \frac{d\sigma^L  (\gamma^* g )}{dx d\eta_1 d\eta_2 dp_T d\bmat{r}_T} = \sigma_0^{g L}\,\sum_{f} H^{L}_{\gamma^*g\to f \bar{f} }(\hat{s},\hat{t}, \hat{u},\mu) \int \frac{d^2\bmat{b}}{(2\pi)^2}  \, \exp(i \bmat{b} \cdot \bmat{r}_T) \, h_1^{\perp}(\xi, \bmat{b}, \mu,\zeta_1) \\
    \times \frac{s_{\bmat {b}}^2 - c_{\bmat {b}}^2}{2}\;S_{\gamma g}(\bmat{b},\zeta_2,\mu) \,\lp \mathcal{C}_{f}(\bmat{b},R,\mu)  J_{f}(p_T,R,\mu) \rp \lp \mathcal{C}_{\bar{f}}(\bmat{b},R,\mu)  J_{\bar{f}}(p_T,R,\mu) \rp\;.   
\end{multline}
We use $s_{\bmat {b}}=\sin\phi_b$ and $c_{\bmat{b} }=\cos\phi_b$ for the sine and cosine of the angle $\phi_b$ between the vectors $\bmat{b}$ and $\bmat{v}_{1T}$, respectively. Each of $d\sigma$ has a hard factor that describes the initiating interaction. The coefficients $\sigma_0^{(f,g), (U,L)}$ are introduced such that the leading order hard functions are normalized to the unity, i.e. $H^{U(L)}_{\text{LO}} = 1 +\mathcal{O}(\alpha_s)$.

The case of heavy hadron pair is very similar. The  measured imbalance $\bmat{r}_T$ is
\begin{equation}
\bmat{r}_T = \bmat{p}_{T}^{H} + \bmat{p}_T^{\bar{H}}\;,
\end{equation}
where the superscript $H$ indicates a generic heavy meson and  $\bar{H}$ the corresponding anti-particle. The imbalance is measured in the Breit frame and assuming the TMD factorization scaling, i.e.,  $|\bmat{r}_T | \ll p_T^{H,\bar H}$. We also assume that  the two heavy mesons are fragmented near the kinematic end-point and carry most of the energy of the heavy quark coming from the hard process. 
The cross-section reads
\begin{multline}
    \frac{d\sigma (\gamma^* g)}{dx d\eta_H d\eta_{\bar{H}} dp_T d\bmat{r}_T} = H^{\mu \nu} _{\gamma^*g\to Q \bar{Q} }(\hat{s},\hat{t}, \hat{u},\mu) \int \frac{d \bmat{b}}{(2\pi)^2}  \, \exp(i \bmat{b} \cdot \bmat{r}_T) \,F_{g, \mu \nu}(\xi, \bmat{b}, \mu,\zeta_1) \\
    \times S_{\gamma g}(\bmat{b},\mu,\zeta_2) \,J_{Q \to H}(\bmat{b}, p_T,m_Q, \mu) \,J_{\bar{Q} \to \bar{H}}(\bmat{b}, p_T, m_Q, \mu)\;.
\end{multline}
 with $\eta_H$ and $\eta_{\bar{H}}$ the pseudo-rapidities of the heavy mesons, $J_{Q \to H}$ the heavy quark jet-functions~\cite{Jaffe:1993ie,Fickinger:2016rfd}. The hard, soft, and beam functions are the same as in the dijet case.
In the hard function we do not consider corrections due to the quark mass and we define
\begin{equation}
p_T  = \frac{|\bmat{p}_{T}^{H} | + |\bmat{p}_{T}^{\bar{H}}| }{2}\;,
\end{equation}

The heavy quark jet functions, $J_{Q\to H}$, can be partially evaluated in perturbation theory as shown in \cite{delCastillo:2020omr}. We work in the limit $p_T\gg m_H\gg\Lambda_{\rm QCD}$ and the heavy quark jet function can be re-factorized  using bHQET.  We also have  $r_T \ll p_T$ so that it is possible to find  large logs of two parametrically different scales  in the fragmentation process,  
\begin{align}\label{eq:hqscales}
\mu_+ = m_Q ,\;\quad  \text{ and } \;\quad  \mu_{\mathcal{J}} = m_Q \frac{r_T}{p_T}\;,
\end{align}
that need to be resummed to ensure the convergence of the expansion.  Following \cite{delCastillo:2020omr} the
jet function can be firstly factorized into a short distance matching coefficient and a bHQET matrix element,
\begin{align}\label{eq:heavyjetfact}
J_{Q \to H} (\bmat{b},p_T,m_Q,\mu) &= H_{+} (m_Q,\mu) \mathcal{J}_{Q \to H} \lp \bmat{b}, \frac{m_Q}{p_T}, \mu \rp,
\end{align}
where the coefficient $H_+$ is
 \begin{equation}
 H_{+} (m_Q,\mu) = | C_{+} (m_Q,\mu) |^2  \;.
\end{equation}
and  the  two-dimensional shape function is defined  in momentum space as
\begin{equation}
\label{eq:hqet-jet}
    \mathcal{J}_{Q \to H} (\bmat{r}) = \frac{1}{ 2\, p^-_H \, N_C } \sum_X \langle 0| \delta^{(2)}\lp \bmat{r} -   i \bmat{v} \,(\bar{v} \cdot \partial \rp)  W_{v}^{\dag} h_{v\beta_+} | X H \rangle \langle X H |  \bar{h}_{v,\beta_+} W_{v}  \, \slashed{\bar v}| 0 \rangle .
\end{equation}
Notice that $\bmat{v}$ is a Euclidean,  two dimensional, transverse component of the light-like four-vector $v^{\mu}$ pointing along the direction of the boosted heavy meson. In position space $ \mathcal{J}_{Q \to H}$ is obtained by  Fourier transformation 
\begin{equation}
\label{eq:FT-jet}
    \mathcal{J}_{Q \to H}\lp \bmat{b}, \frac{m_Q}{p_T}, \mu \rp = \int d\bmat{r} \exp(i \bmat{b} \cdot \bmat{r}) \mathcal{J}_{Q \to H}(\bmat{r})\;.
\end{equation}
The one-loop expression for these quantities are calculated in  \cite{delCastillo:2020omr}.
\section{Cross-sections used in phenomenology}
\label{sec:3}
The cross-sections presented in previous section are usually partially integrated in phenomenological observables. We discuss here these integrations, which also allow us to relate the normalization of our cross-section with the ones obtained in the literature.

\subsection{Extracting the Born-level cross-sections}
The tree level cross-sections for the dijet and hadron pair production were considered at tree level in ref.~\cite{Boer:2016fqd}. We start considering the gluon case,  from which one can easily deduce also the quark case. The gluon  hard contribution to the cross-section is described by 
\begin{equation}
    \frac{d\sigma(\gamma^* g )}{dx d\eta_1 d\eta_2 d\bmat{p}_T d\bmat{r}_T} =  \frac{\mathcal{N}}{x s} \lb A_0 + A_1\cos 2\phi'_{p} +\cdots +B_0 \cos 2\phi'_r + \cdots \rb ,
\end{equation}
and the azimuthal angles ($\phi'_r$, $\phi'_p$) of vectors  $\bmat{p}_T ,\bmat{r}_T$ are measured with respect to the lepton plane. However, our preferred frame is the one where the   $\phi_\ell$  angle is measured in the plane defined by $\bmat{p}_T$ and $\bmat{q}_T$, the sum of the lepton momenta, and   $\phi_r$ is the azimuthal angle between $\bmat{r}_T$ and $\bmat{p}_T$.
In this frame and integrating over the angle  $\phi_\ell$ we are left with: 
\begin{equation}
    \frac{d\sigma  (\gamma^* g )}{dx d\eta_1 d\eta_2 dp_T d\bmat{r}_T } =  2 \pi p_T \frac{\mathcal{N}}{x s} \lb A_0  + B_2 \cos (2\phi_r) \rb ,
\end{equation}
with the factor $2\pi$ coming from $\phi_\ell$ integration.
 The LO expressions  
 are obtained by separating the unpolarized  and linearly polarized gluon contributions and Fourier transforming. 
 The unpolarized partonic part has a similar form also for quarks, so that
  we find 
\begin{eqnarray}\label{eq:FactGamGU_LO}
    \frac{d\sigma^U  (\gamma^* g )}{dx d\eta_1 d\eta_2 dp_T d\bmat{r}_T} \Bvert_{\text{LO}} = 
    \sigma_0^{g U}  \, \int \frac{d^2\bmat{b}}{(2\pi)^2}  \, \exp(i \bmat{b} \cdot \bmat{r}_T) \, f_1^g(\xi, \bmat{b})  = \sigma_0^{g U}  f_1^g(\xi, \bmat{r}_T)  \;,    
\\
\label{eq:FactGamfU_LO}
    \frac{d\sigma^U  (\gamma^* f )}{dx d\eta_1 d\eta_2 dp_T d\bmat{r}_T} \Bvert_{\text{LO}} = \sigma_0^{f U}  \, \int \frac{d^2\bmat{b}}{(2\pi)^2}  \, \exp(i \bmat{b} \cdot \bmat{r}_T) \, f_1^{f}(\xi, \bmat{b})  = \sigma_0^{f U}  f_1^{f}(\xi, \bmat{r}_T)  \;,    
\end{eqnarray}
The same for the linearly polarized gluons gives 
\begin{align}\label{eq:FactGamGL_LO}
    \frac{d\sigma^L  (\gamma^* g )}{dx d\eta_1 d\eta_2 dp_T d\bmat{r}_T} \Bvert_{\text{LO}} &= \sigma_0^{g L}\, \int \frac{d^2\bmat{b}}{(2\pi)^2}  \, \exp(i \bmat{r}_T \cdot \bmat{b}) \,  \frac{\sin^2\phi_b - \cos^2 \phi_b}{2}\;h_1^{\perp}(\xi, \bmat{b}) \;.   \nl
    & = -\sigma_0^{g L}\, \int \frac{b\, db \, d\phi_b}{8\pi^2}  \exp \lp i r_T b \cos (\phi_b -\phi_r) \rp \;\cos(2\phi_b) \,h_1^{\perp}(\xi, \bmat{b}) \nl
    & = \cos(2\phi_r)\;\sigma_0^{g L} \;\int \frac{b\, db }{4\pi} J_2( r_T b) \,h_1^{\perp}(\xi, \bmat{b}) \nl
    & = -\frac{\cos(2\phi_r)}{2} \;\sigma_0^{g L}\,h_1^{\perp}(\xi, \bmat{r}_T),
\end{align}
where notice that $h_1^{\perp}(\xi,r_T)$ is not the direct Fourier transform of $h_1^{\perp}(\xi,b)$ and both functions can be related through eq.~(2.20) in \cite{Echevarria:2015uaa}. We obtain the $\sigma_0^{(g,f) (U,L)}$ prefactors from the structure functions given in eqs.~(3.3, 3.5) in \cite{Boer:2016fqd} and we list them in appendix \ref{app:Prefactors}.
\subsection{Angle integrated and azimuthally modulated cross-section }
The scalar cross-section that we finally consider in the phenomenological studies is obtained by integrating over the $\phi_r$ angle
\begin{equation}
    \frac{d\sigma}{d\Pi dr_T} = r_T \int_{-\pi}^{+\pi} d\phi_r \frac{d\sigma}{d\Pi d\bmat{r}_T}, 
\end{equation}
where $d\Pi =dx d\eta_1 d\eta_2 dp_T$.
Because the factorized cross-section is always expressed in position space one can write (here $J_{0,2}$ are Bessel functions)
\begin{align}
    \frac{d\sigma}{d\Pi d r_T} &= r_T \int_{-\pi}^{+\pi} d\phi_r \int \frac{d\bmat{b} }{(2\pi)^2} \exp\lb i r_T b \cos(\phi_b-\phi_r) \rb \; \frac{d \tilde\sigma(\bmat{b})}{d\Pi d \bmat{b}} \nl
    & = r_T \int_{0}^{\infty} \frac{b\,db}{2 \pi } J_0(r_T b)  \int_{-\pi}^{+\pi} d\phi_b\,
    \frac{d \tilde \sigma(\bmat{b})}{d\Pi d \bmat{b}}
    \nl
    & = r_T \int_{0}^{\infty} \frac{b\,db}{2 \pi } J_0(r_T b)  \int_{-\pi}^{+\pi} d\phi_b\, \lb 
    \frac{d \tilde\sigma^U(\bmat{b})}{d\Pi d \bmat{b}}
    -\frac{\cos 2\phi_b}{2} \frac{d \tilde\sigma^L(\bmat{b})}{d\Pi d \bmat{b}} \rb , \label{eq:ds1}
\end{align}
where $d \sigma^U=d \sigma^U(\gamma^* f) +d \sigma^U(\gamma^* g)$, $d \sigma^L=d \sigma^L(\gamma^* g)$ for the dijet case and $d \sigma^{U,L}=d \sigma^{U,L}(\gamma^* g)$ for the heavy hadron pair case.

In our phenomenological analysis we consider also the azimuthal angle average
\begin{equation}\label{eq:ds2}
    \langle \cos 2\phi_r \rangle \equiv 
    \Bigg[
    \int_{-\pi}^{+\pi} d\phi_r \cos 2 \phi_r \frac{d\sigma}{d\Pi d\bmat{r}_T} 
    \Bigg]
    \Bigg{/}  
    \frac{d\sigma}{d\Pi dr_T}.
\end{equation}
The denominator is what we have discussed in the previous section, so now we will only focus on the numerator, 
\begin{align}
    \int_{-\pi}^{+\pi} d\phi_r \cos 2 \phi_r \frac{d\sigma}{d\Pi d\bmat{r}_T} &= r_T \int_{-\pi}^{+\pi} d\phi_r \cos 2\phi_r \int \frac{d\bmat{b}}{(2\pi)^2} \exp \lb i r_T b \cos(\phi_b -\phi_r) \rb  \nonumber 
    \frac{d \tilde\sigma(\bmat{b})}{d\Pi d \bmat{b}} \nl
    & = r_T \int_{0}^{\infty} \frac{b\,db}{2\pi} J_2(r_T b) \int_{-\pi}^{+\pi} d\phi_b  \cos 2\phi_b \; 
    \frac{d\tilde \sigma(\bmat{b})}{d\Pi  d \bmat{b}}
     \nl
    &=    r_T \int_{0}^{\infty} \frac{b\,db}{2\pi} J_2(r_T b)   \int_{-\pi}^{+\pi} d\phi_b\, 
    \lb  \cos 2\phi_b\; 
    \frac{d \tilde\sigma^U(\bmat{b})}{d\Pi d \bmat{b}}
    -\frac{\cos^2 2\phi_b}{2} \frac{d \tilde\sigma^L(\bmat{b})}{d\Pi  d \bmat{b}} \rb .\label{eq:ds3}
\end{align}
Eqs.~(\ref{eq:ds1} - \ref{eq:ds3}) show the relation among the final cross-section in momentum space and the factorized cross-sections in position space, where we have  distinguished between the unpolarized cross-sections  generated by gluons and quarks, $d\tilde  \sigma^U=d\tilde  \sigma^U_g +d\tilde  \sigma^U_f$ and the contribution from linearly polarized gluons $d\tilde  \sigma^L$.
\section{Evolution kernels with angular dependent anomalous dimensions}
\label{sec:eka}
The anomalous dimensions of soft, collinear-soft and heavy-jets matrix elements
are  angular dependent and  complex valued, and both these features are not common in literature.    The angular dependence  is parameterized here by the angle $\phi_b$. In order to  discuss the issue we show in sec.~\ref{subsec:soft} the dijet soft function  as obtained in \cite{delCastillo:2020omr} and then we extend the conclusions to the other functions that have a similar angular dependent structure.  The angular dependence of the anomalous dimension is strictly correlated with the imaginary parts of the cross-section in position space. The treatment of the angular dependence in the evolution has been discussed also in \cite{Kang:2020xez,Sun:2014gfa,Sun:2015doa} which propose an approximate treatment. In sec.~\ref{sec:ang} we perform  an original analysis discussing  in detail how passing from position space to momentum space  in the cross-section allows to obtain a real valued cross-section, including resummation. The proof of this statement is here provided at one-loop, and the same mechanism can be conjectured to work at all orders in perturbation theory.

\subsection{Dijet soft function and angle dependent anomalous dimensions}
\label{subsec:soft}
The dijet soft function presents several properties that result particularly interesting.
The renormalized soft function is obtained from its bare expression
\begin{equation}
 S_{\gamma i}^{\text{bare}}(\bmat{b}, \zeta_2) = Z_{\gamma i}^{S}(\bmat{b}, \mu, \zeta_2)  S_{\gamma i}(\bmat{b}, \mu, \zeta_2),
 \end{equation}
 where the factor $Z_{\gamma i}^{S}$ was calculated in \cite{delCastillo:2020omr}. For our purposes it is sufficient to  report expression of the soft function 
in the $\overline{\text{MS}}$ scheme  for the $(\gamma^* g)$-channel  
\begin{multline}
S_{\gamma g} (\bmat{b},\mu,\zeta_2) =   1 + a_s\lbc 
 C_F \lb \frac{\pi^2}{3}  + 2 \ln^2 \lp \frac{B \mu^2 e^{2\gamma_E}}{ -A_{\bmat{b}}}\rp + 4 \text{Li}_2(1+A_{\bmat{b}})  \rb \\[5pt]
 +C_A \lb  - 2 \ln (B \mu^2 e^{2\gamma_E})   \ln \zeta_2  - \ln^2(-A_{\bmat{b}}) - \frac{\pi^2}{3} - 2  \text{Li}_2(1+A_{\bmat{b}}) \rb 
\rbc +{\cal O}(a_s^2),
\end{multline}
and for the ($\gamma^* f$)-channel 
\begin{multline}
 S_{\gamma f} (\bmat{b},\mu,\zeta_2) = 1 + a_s \lbc
C_A \lb  \frac{\pi^2}{6}  +  \ln^2 \lp \frac{B \mu^2 e^{2\gamma_E}}{ -A_{\bmat{b}}}\rp + 2 \text{Li}_2(1+A_{\bmat{b}}) + 2\ln(B \mu^2 e^{2\gamma_E} )   \ln \frac{(n \cdot v_1) (\bmat{v}_2 \cdot \bmat{b})}{(n \cdot v_2) (\bmat{v}_1 \cdot \bmat{b})}   \rb \\[5pt]
+ C_F \ln(B \mu^2 e^{2\gamma_E} ) \lb \ln (B \mu^2 e^{2\gamma_E}) -2 \ln \zeta_2 + 2 \ln \lp\frac{2  (n \cdot v_2)}{(v_1\cdot v_2) (n \cdot v_2)} \rp  + 4\,\text{ln} (-i \,\bmat{v}_1 \cdot \hat{\bmat{b}}) 
\rb - C_F \frac{\pi^2}{6} 
 \rbc +{\cal O}(a_s^2),
\end{multline}
respectively, where we have $A_{{\bmat{b}}} = \frac{(v_1 \cdot v_2)}{2 \,(v_{1} \cdot \hat{b})   \,(v_{2} \cdot \hat{b})  }$. The expressions above depend explicitly on the angle $\phi_b$  and we have used $ \hat{\bmat{b}}=\bmat{b}/|\bmat{b}|$. The rapidity scale $\zeta_2$ is responsible for the rapidity evolution of this factor and it is related to the TMD rapidity scale $\zeta_1$ by the consistency constraint  
\begin{equation}
\label{eq:zz}
\zeta_1 \, \zeta_2 = \frac{(k^-)^2}{A_n} =  \frac{\hat{u}  \;\hat{t}}{\hat{s}} .
\end{equation}
The values of $\zeta_1$ and $\zeta_2$ that minimize hard logs are
\begin{align}
\zeta_1 =p_T^2,\;\quad \zeta_2  =1.
\end{align}
Despite the fact that the scale $\zeta_2$ is dimensionless there are some formal similarities in the evolution of this soft function and the TMD. 
The dijet soft function double scale evolution is dictated by 
\begin{align}
    \frac{d}{d \ln \mu} S_{\gamma i} (\bmat{b}, \zeta, \mu) & = \gamma_{S_{\gamma i}}(\bmat{b},\mu,\zeta) S_{\gamma i}(\bmat{b},\zeta,\mu), \label{eq:AD_evol} \\
    \frac{d}{d \ln \zeta} S_{\gamma i} (\bmat{b}, \zeta, \mu) & = -\mathcal{D}_{ i}(\mu,b) S_{\gamma i}(\bmat{b},\zeta,\mu), \label{eq:RAD_evol}
\end{align}
where $i$ is gluon or quark. The rapidity evolution kernel $\mathcal{D}_{ i} $ is the same as in the TMD case, while for the rest we need a special treatment.
\subsection{Treatment of angular dependent anomalous dimensions and resummation}
\label{sec:ang}
The dijet soft function, the collinear-soft function in dijets and the heavy meson jet function in hadron pair production have, as a common feature, an anomalous dimension that is $\phi_b$-dependent and that can include some imaginary parts.
The general structure of the anomalous dimension  for these cases is
\begin{equation}
    \gamma_i(\bmat{b},\mu) =  \gcusp[\alpha_s] \,\lp  c_i \,2 \ln |\cos\phi_b| - c'_i\,i \pi \Theta(\phi_b) \rp + \text{other $\phi_b$ independent terms}
\end{equation}
where 
\begin{equation}
    \Theta(\phi_b) = 
    \begin{cases}
    +1  &:  -\pi/2 < \phi_b < \pi/2 \\
    -1 &: \text{otherwise}
    \end{cases}
\end{equation}
and 
the consistency of anomalous dimensions requires
\begin{equation}\label{eq:c}
    \sum_i c_i=  \sum_i c'_i = 0 ,
\end{equation}
where $c_i$ and $c_i'$ are the color coefficients that multiply the corresponding part of the cusp anomalous dimension. The construction of the evolution kernel involves an integration over the angle  $\phi_b$ and after integration all imaginary parts cancel consistently. Note that this does not mean that one can ignore the imaginary components of the anomalous dimensions or fixed order functions as contributions of imaginary terms yield real contributions to the final result.

In order to show the cancellation of the imaginary part we separate the angular dependent part of the evolution kernel from the rest.  For instance for the soft function we have
\begin{align}
    S_{\gamma i}\left(\bmat b, \mu_f, \zeta_{2,f}\right) & = \exp \left[\int_{P} \lp 
    \gamma_{S_{\gamma i}}(\bmat b, \mu, \zeta_2) d \ln \mu -\mathcal{D}_i(\mu, b) d \ln \zeta_2 \rp \right] S\left(\bmat b ; \mu_{0}, \zeta_{2,0}\right) \nl
    & = \exp \left[\int_{P} \lp \gamma^{\phi}_{S_{\gamma i}}(\phi) d \ln \mu + \bar{\gamma}_{S_{\gamma i}}(b, \mu, \zeta_2) d \ln \mu -\mathcal{D}_i(\mu, b) d \ln \zeta_2 \rp \right] S_{\gamma i}\left(\bmat b , \mu_{0}, \zeta_{2,0}\right)\nl
    & = \exp \left[\int_{\mu_0}^{\mu_f} \lp \gamma^{\phi}_{S_{\gamma i}}(\phi) d \ln \mu \rp \right] \exp \left[ \int_P \lp \bar{\gamma}_{S_{\gamma i}}(b, \mu, \zeta_2) d \ln \mu -\mathcal{D}_i(\mu, b) d \ln \zeta_2 \rp \right] S_{\gamma i}\left(\bmat b , \mu_{0}, \zeta_{2,0}\right) \nl
    & = \mathcal{R}_{S_{\gamma i}}^{\phi}\ \mathcal{R}_{S_{\gamma i}}\ S_{\gamma i}\left(\bmat b , \mu_{0}, \zeta_{2,0}\right).
\end{align}
The evolution factor is so splitted  in an angular dependent part and the rest. The splitting is clearly not unique, however this should not affect the final result once the angular integration is performed.
We have
\begin{align}
   \frac{d}{d\ln \mu} \mathcal{R}_{S_{\gamma i}}^{\phi}&=\gamma^{\phi}_{S}(\phi) \mathcal{R}_{S_{\gamma i}}^{\phi}, \quad
   \frac{d}{d\ln \zeta} \mathcal{R}_{S_{\gamma i}}^{\phi}= 0 \\
   \label{eq:kernelS}
     \frac{d}{d\ln \mu} \mathcal{R}_{S_{\gamma i}}&=\bar \gamma_{S} \mathcal{R}_{S_{\gamma i}}, \quad
   \frac{d}{d\ln \zeta} \mathcal{R}_{S_{\gamma i}}= -{\mathcal{D}_i} \mathcal{R}_{S_{\gamma i}} ,
\end{align}
A similar splitting is done for all other other angular dependent evolution kernels, for which we have used
\begin{align}
    \gamma_{S_{\gamma g}}^{\phi} & = \gamma_{\mathrm{cusp}} \lb 4 C_F \ln | \cos \phi_b | \rb, \\
    \gamma_{S_{\gamma f}}^{\phi} & = \gamma_{\mathrm{cusp}} \lb 2 (C_F + C_A) \ln | \cos \phi_b | - (C_F - C_A) i \pi \Theta (\phi_b) \rb, \\
    \gamma_{\mathcal{C}_{g}}^{\phi} & = \gamma_{\mathrm{cusp}} C_A \lb - 2 \ln | \cos \phi_b | - i \pi \Theta (\phi_b) \rb, \\
    \gamma_{\mathcal{C}_{f}}^{\phi} & = \gamma_{\mathrm{cusp}} C_F \lb - 2 \ln | \cos \phi_b | \pm i \pi \Theta (\phi_b) \rb, \\
    \gamma_J^{\phi} & = \gamma_{\mathrm{cusp}} C_F \lb -2  \ln | \cos \phi_b | \pm i \pi \Theta (\phi_b) \rb,
\end{align}
where the $\pm$ sign refers to quark and anti-quark jet respectively.
The angular independent part of the anomalous dimension is simply obtained from  the complete expression of the anomalous dimension
\begin{align}
   \gamma_{i}=\gamma_{i}^{\phi}+\bar \gamma_{i} 
\end{align}
with $i=S_{\gamma g},S_{\gamma f},{\cal C}_g,{\cal C}_f, J$ for each channel. The complete list of one loop anomalous dimensions can be found in appendix \ref{app:AD}.

In order to describe the implications of the angle integration we  consider here the dijet production case, being the HHP one very similar.
The separation of the evolution kernels of all functions in an angular dependent part and the rest allows to write the  resummed cross-section in  position space as 
\begin{multline}\label{eq:NNLL_resum_XS}
d\tilde \sigma (\bmat{b}) \sim \exp \lb   \mathcal{A}(\{\mu_i\})\,  2 \ln |\cos\phi_b| -  \mathcal{B}(\{\mu_i\})\,i \pi \Theta(\phi_b) \rb  \mathcal{R}(\{\mu_k\} \to \mu) \\[5pt]
\times\lb  1+ \sum_{k\in\{H,F,J,S,\mathcal{C}\}} a_s(\mu_k) f_k^{[1]}(b,\cos \phi_b)\rb\\
= |\cos\phi_b|^{2 \mathcal{A}} \lp \cos( \mathcal{B} \pi) - i \Theta(\phi_b)\sin( \mathcal{B} \pi) \rp \mathcal{R}(\{\mu_k\} \to \mu)
\lb  1+ \sum_{k\in\{H,F,J,S,\mathcal{C}\}} a_s(\mu_k) f_k^{[1]}(b,\cos \phi_b)\rb
\end{multline}
where we have omitted global scale independent hard factors and non-perturbative contributions that we assume  independent of the angle $\phi_b$. In this equation we have combined the evolution kernel of the functions that appear in the cross-section in
\begin{align}\label{eq:A_def}
    \mathcal{A}(\{\mu_i\}) &= \sum_{i\in \{S,\mathcal{C}\}} c_i \,\int_{\mu_i}^{\mu} \gcusp[\alpha_s] d\ln \mu' \,,&\mathcal{B}(\{\mu_i\}) &= \sum_{i\in \{S,\mathcal{C}\}} c'_i \,\int_{\mu_i}^{\mu} \gcusp[\alpha_s] d\ln \mu' \,
\end{align}
which are independent of the factorization scale $\mu$ because of eq.~(\ref{eq:c}), and an angle independent part $\mathcal{R}(\{\mu_k\} \to \mu)$. The perturbative parts of all functions that appear in the cross-sections at one loop are collected in the factor $
\lb  1+ \sum_{k\in\{H,F,J,S,\mathcal{C}\}} a_s(\mu_k) f_k^{[1]}\rb$.
In eq.~(\ref{eq:NNLL_resum_XS}) the imaginary part of the cross-section is proportional to the odd function $\Theta(\phi_b)$. We expect so that this part cancel at all orders in perturbation theory when the $\phi_b$ the Fourier transform is performed like in eq.~(\ref{eq:ds1}-\ref{eq:ds3}).
In the next subsection we show that this is explicitly the case at one loop.

In order to organize the discussion we firstly consider the case of  $d\tilde{\sigma}^{U}$ and then deduce the necessary modifications to get the Fourier transform of  $d\tilde{\sigma}^{L}$. The angular integration  of the cross-section at one loop can always be expressed in terms of
 the following basic integrals:
\begin{equation}
  I_n (\mathcal{A})\equiv  \int_{-\pi}^{+\pi} d\phi_b\,  |\cos\phi_b|^{2 \mathcal{A}} \; \ln^{n} |\cos\phi_b|
\end{equation}
where 
\begin{align}\label{eq:integral_basis}
     I_0(\mathcal{A}) &= \frac{2 \sqrt{\pi} \;\Gamma (1/2+\mathcal{A} )}{\Gamma(1+\mathcal{A})},   \nl
     I_1(\mathcal{A}) &= \frac{\sqrt{\pi} \;\Gamma(1/2+\mathcal{A})}{\Gamma(1+\mathcal{A})} \; (H_{\mathcal{A}-1/2} - H_{\mathcal{A}}) \nl
     I_2(\mathcal{A}) &= \frac{\sqrt{\pi} \;\Gamma(1/2+\mathcal{A})}{2 \Gamma(1+\mathcal{A})} \; \lb (H_{\mathcal{A}-1/2} - H_{\mathcal{A}})^2 +\psi^{(1)}\lp\frac{1}{2} + \mathcal{A}\rp - \psi^{(1)}(1+\mathcal{A}) \rb  .
\end{align}
The results on the r.h.s. of eq.~(\ref{eq:integral_basis}) are defined only for $\mathcal{A} > -1/2$. In order to satisfy this condition  for every value of $b$ the scales $\mu_S$ and $\mu_{\mathcal{C}}$ must be chosen appropriately, which at the current  level of precision can be done. Next  we show the contribution of  each term of $f_k^{[1]}$ to the cross-section:
\begin{itemize}
    \item \emph{Constant terms:} In this case we consider the leading order term in eq.~(\ref{eq:NNLL_resum_XS}) together with all terms in the functions $f_{k}^{[1]}$ that are $\cos \phi_b$ independent. The integral we need to perform is, 
    \begin{align}
        I_{\text{const.}}(\mathcal{A},\mathcal{B}) \equiv \int_{-\pi}^{+\pi} d\phi_b\,  |\cos\phi_b|^{2 \mathcal{A}} \lp \cos( \mathcal{B} \pi) - i \Theta(\phi_b)\sin( \mathcal{B} \pi) \rp = I_{0}(\mathcal{A}) \cos( \mathcal{B} \pi) 
    \end{align}
    and the imaginary part cancels because of parity.
    \item \emph{Single logarithmic terms:} In this case we consider terms proportional to $\ln (-i \cos\phi_b)$ that appear in the soft and collinear-soft terms $f_{S}^{[1]}$ and $f_{\mathcal{C}}^{[1]}$ respectively. The relevant integral that we need to perform is 
    \begin{equation}
         I_{\text{log}}(\mathcal{A},\mathcal{B}) \equiv \int_{-\pi}^{+\pi} d\phi_b\,  |\cos\phi_b|^{2 \mathcal{A}} \lp \cos( \mathcal{B} \pi) - i \Theta(\phi_b)\sin( \mathcal{B} \pi) \rp\; \ln (-i \cos\phi_b).
    \end{equation}
    Expanding the logarithm as follows 
    \begin{equation}
         \ln (-i \cos\phi_b) = \ln |\cos \phi_b | -\frac{i \pi}{2} \Theta (\phi_b)
    \end{equation}
    we can split the main integral into  one part proportional to $\cos( \mathcal{B} \pi)$ and a second part proportional to $\sin(\mathcal{B} \pi)$. In terms of the integrals in \eq{integral_basis} we obtain the real valued result
    \begin{equation}
         I_{\text{log}}(\mathcal{A},\mathcal{B}) = I_1 (\mathcal{A}) \cos( \mathcal{B} \pi) - \frac{\pi}{2} I_0 (\mathcal{A}) \sin( \mathcal{B} \pi) .
    \end{equation}
    \item \emph{Logarithmic $A_{\bmat{b}}$ terms:} In the soft function we observe the presence of terms that involve the following combination,
    \begin{equation}
        A_{\bmat{b}} = \frac{(v_1 \cdot v_2)}{2(v_1 \cdot \hat{b})(v_1 \cdot \hat{b})} = -\frac{\hat{s}}{4 p_T^2 (\cos \phi_b)^2}
    \end{equation}
    Particularly this term appears in logarithms and di-logarithms. Here we address the double logarithmic terms, i.e.,
    \begin{equation}
        \ln^2(-A_{\bmat{b}}) = \ln^2\lp \frac{\hat{s}}{4p_T^2} \rp - 4 \ln \lp \frac{\hat{s}}{4p_T^2} \rp \ln |\cos \phi_b | + 4 \ln^2 |\cos \phi_b |  
    \end{equation}
    Integrating over $\phi_b$ the above we obtain three contributions: a constant term, a single-log and a double-log. Thus we can immediately write,
    \begin{align}
        I_{\text{logA}}(\mathcal{A},\mathcal{B}) &\equiv \int_{-\pi}^{+\pi} d\phi_b\,  |\cos\phi_b|^{2 \mathcal{A}} \lp \cos( \mathcal{B} \pi) - i \Theta(\phi_b)\sin( \mathcal{B} \pi) \rp\; \ln^2 (-A_{\bmat{b}}) \nl
        & = \lb \ln^2\lp \frac{\hat{s}}{4p_T^2} \rp  I_{0}(\mathcal{A})  - 4 \ln \lp \frac{\hat{s}}{4p_T^2} \rp I_{1}(\mathcal{A})  + 4  I_{2}(\mathcal{A}) \rb \cos( \mathcal{B} \pi) .
    \end{align}
    
    \item \emph{Double logarithmic terms:} In this case we consider terms proportional to $\ln ^2(-i \cos\phi_b)$ that appear in the soft and collinear-soft functions. The relevant integral that we need to perform is 
    \begin{equation}
        I_{\text{log2}}(\mathcal{A},\mathcal{B}) \equiv \int_{-\pi}^{+\pi} d\phi_b\,  |\cos\phi_b|^{2 \mathcal{A}} \lp \cos( \mathcal{B} \pi) - i \Theta(\phi_b)\sin( \mathcal{B} \pi) \rp\; \ln^2 (-i \cos\phi_b)
    \end{equation}
    we can then expand the logarithm as follows 
    \begin{equation}
         \ln^2 (-i \cos\phi_b) = \ln^2 |\cos \phi_b | -\frac{\pi^2}{4} -i \pi \Theta (\phi_b) \ln |\cos \phi_b |
    \end{equation}
    then we can expand as before in terms proportional to $\sin( \mathcal{B} \pi)$ and  terms proportional to $\cos( \mathcal{B} \pi)$
    \begin{equation}
         I_{\text{log2}}(\mathcal{A},\mathcal{B}) = \lb -\frac{\pi^2}{4} I_0(\mathcal{A})+  I_2 (\mathcal{A}) \rb \cos( \mathcal{B} \pi) -\pi I_{1}(\mathcal{A}) \sin( \mathcal{B} \pi).
    \end{equation}
\end{itemize}
In addition we have poly-logarithmic terms which are proportional to $\text{Li}_2(1-A_{\bmat{b}})$,
\begin{equation}
    I_{\text{Li}}(\mathcal{A},\mathcal{B}, \bar{c})\equiv \cos( \mathcal{B} \pi)  \int_{-\pi}^{+\pi} d\phi_b\,  |\cos\phi_b|^{2 \mathcal{A}} \; \text{Li}_2 \lp1 - \frac{1}{\bar{c}\; \cos^2\phi_b}\rp,
\end{equation}
where 
\begin{align}
    \bar{c} &\equiv \frac{4p_T^2}{\hat{s}}\;,& 0&\,<\,\bar{c}\,\leq\,1,
\end{align}
and where we have dropped the terms proportional to $\sin(\mathcal{B} \pi)$ since they vanish after integration from parity. We can then use properties of the poly-logarithm in order to express this integral in the following form, 
\begin{multline}
     I_{\text{Li}}(\mathcal{A},\mathcal{B}, \bar{c} ) = \cos(\mathcal{B} \pi) \lbc -\frac{1}{2} \lb \lp \ln^2\bar{c} +\frac{\pi^2}{3}  \rp I_0(\mathcal{A}) + 4 \ln \bar{c} I_1(\mathcal{A}) + 4 I_2(\mathcal{A})  \rb  \\[5pt]
     +\int_{-\pi}^{+\pi} d\phi_b\,  |\cos\phi_b|^{2 \mathcal{A}} \; \lb \text{Li}_2 (\bar{c} \,\cos^2\phi_b) +\ln(\bar{c} \,\cos^2\phi_b) \ln(1-\bar{c} \,\cos^2\phi_b) \rb \rbc
\end{multline}
We can then expand the poly-logarithms and the $\ln(1-\bar{c} \,\cos^2\phi_b)$ in the region $0<c\leq1$. This allows us to perform the integral over $\phi_b$ and obtain the following,
\begin{multline}
     I_{\text{Li}}(\mathcal{A},\mathcal{B}, \bar{c} ) = \cos(\mathcal{B} \pi)  \lbc -\frac{1}{2} \lb \lp \ln^2\bar{c} +\frac{\pi^2}{3}  \rp I_0(\mathcal{A}) + 4 I_1(\mathcal{A})\, \ln \bar{c}  + 4 I_2(\mathcal{A})  \rb  \\[5pt]
     -\sum_{n=1}^{\infty} \frac{{\bar{c}}^{\,n}}{n} \lb \lp \ln \bar{c} -\frac{1}{n} \rp I_0(\mathcal{A}+n) + 2 I_1(\mathcal{A}+n)\rb \rbc 
\end{multline}
Although this expression is exact it includes a  sum extending to infinity. In numerical applications one can truncate this sum at the desired value to achieve a certain numerical precision.
This concludes the discussion of all possible cases that appear in the Fourier transform of the unpolarized cross-section.

The part of the cross-section relative to  linearly polarized gluons can be treated similarly. In this case we need to incorporate an additional $\cos 2\phi_b$ term in the integrals but this is the only change since the soft and collinear-soft functions that appear in the two contributions are the same  as for the dijet case.
Using the trigonometric identity 
$
    \cos 2\phi_b = 2 \cos^2 \phi_b -1
$ 
 the integrals for  this case can be deduced from the discussion of the unpolarized cross-section by the replacement 
\begin{equation}
    I_n(\mathcal{A}) \quad \longrightarrow \quad - I_n(\mathcal{A}+1) + \frac{1}{2}I_n(\mathcal{A})\;.
\end{equation} 
Equivalently for the case of angular modulation in eq.~(\ref{eq:ds2}) the following transformations have to be performed,  
\begin{align}
   d\tilde\sigma^{U}(\bmat{b})&: \quad\quad I_n(\mathcal{A}) \quad  \longrightarrow \quad  - I_n(\mathcal{A}) +2 I_n(\mathcal{A}+1)  \nl
   d\tilde\sigma^{L}(\bmat{b})&:\quad\quad I_n(\mathcal{A}) \quad  \longrightarrow \quad  -I_n(\mathcal{A})+2 I_n(\mathcal{A}+1) -2 I_n(\mathcal{A}+2)  \;.
\end{align}
In all these cases one obtains a cancellation of the imaginary part of the cross-section.

Treating perturbatively the angular integration as discussed in this section leads to write   
 eq.~(\ref{eq:ds1}) for the dijet case as 
\begin{align}\label{eq:siJ}
    \frac{d\sigma}{d\Pi dr_T}&=\frac{d\sigma^U(\gamma^* g)}{d\Pi dr_T}+\frac{d\sigma^U(\gamma^* f)}{d\Pi dr_T}+\frac{d\sigma^L(\gamma^* g)}{d\Pi dr_T},
\end{align}
where
\begin{align}
    \label{eq:sifg1}
    \frac{d \sigma^{U}(\gamma^*g)}{d \Pi\,d r_T} &= \sum_f \sigma_0^{gU}H^{U}_{\gamma^*g\to f \bar{f} }(\hat{s},\hat{t}, \hat{u},\mu=p_T) J_{f}(p_T,R,\mu_J) J_{\bar f}(p_T,R,\mu_J) \nonumber\\
    &\times \int_0^{+\infty} b db\, J_0(b r_T) f_1^g(\xi, \bmat{b}) \mathcal{R}_g \lp (\{\mu_k\},\zeta_{1,0},\zeta_{2,0}) \to (p_T,p_T^2,1) \rp  \hat \sigma^{U}_g (b, R, \{\mu_i\}),
    \\ \label{eq:siff1}
    \frac{d \sigma^{U}(\gamma^*f)}{d \Pi\,d r_T} &= \sum_{f,\bar f} \sigma_0^{fU}H^{U}_{\gamma^*f\to g f }(\hat{s},\hat{t}, \hat{u},\mu=p_T) J_{f}(p_T,R,\mu_J) J_{g}(p_T,R,\mu_J) \nonumber\\ 
    &\times \int_0^{+\infty} b db\, J_0(b r_T) f_1^f(\xi, \bmat{b}) \mathcal{R}_q \lp (\{\mu_k\},\zeta_{1,0},\zeta_{2,0}) \to (p_T,p_T^2,1) \rp  \hat\sigma^{U}_f(b, R, \{\mu_i\}),
    \\ \label{eq:sifg1L}
    \frac{d \sigma^{L}(\gamma^*g)}{d \Pi\,d r_T} &= \sum_f \sigma_0^{gL}H^{L}_{\gamma^*g\to f \bar{f} }(\hat{s},\hat{t}, \hat{u},\mu=p_T) J_{f}(p_T,R,\mu_J) J_{\bar f}(p_T,R,\mu_J) \nonumber\\
    &\times \int_0^{+\infty} b db\, J_0(b r_T) h_1^\perp(\xi, \bmat{b}) \mathcal{R}_g \lp (\{\mu_k\},\zeta_{1,0},\zeta_{2,0}) \to (p_T,p_T^2,1) \rp  \hat \sigma^{L}_g (b, R, \{\mu_i\}),
\end{align}
where $\mathcal{R}_{f,g}$ are products of evolution kernels to be described in the next section, 
and $\hat\sigma^{U,L}_{f,g}$ are the result of $\phi_b$ angular integration and can be written as
\begin{align}\label{eq:hats1}
    \hat\sigma^{U}_g &= I_\mathrm{const.}^{gU} + a_s(\mu_{\mathcal{C}})\mathcal{C}_f^{U}(b,R,\mu_{\mathcal{C}}) + a_s(\mu_{\mathcal{C}})\mathcal{C}_{\bar f}^{U}(b,R,\mu_{\mathcal{C}}) + a_s(\mu_0) S_{\gamma g}^{U}(b,\zeta_2,\mu_0),\\  \label{eq:hats2}
    \hat\sigma^{U}_f &= I_\mathrm{const.}^{fU} + a_s(\mu_{\mathcal{C}})\mathcal{C}_f^{U}(b,R,\mu_{\mathcal{C}}) + a_s(\mu_{\mathcal{C}})\mathcal{C}_{g}^{U}(b,R,\mu_{\mathcal{C}}) + a_s(\mu_0) S_{\gamma f}^{U}(b,\zeta_2,\mu_0),\\ \label{eq:hats3}
     \hat\sigma^{L}_g &= I_\mathrm{const.}^{gL} + a_s(\mu_{\mathcal{C}})\mathcal{C}_f^{L}(b,R,\mu_{\mathcal{C}}) + a_s(\mu_{\mathcal{C}})\mathcal{C}_{\bar f}^{L}(b,R,\mu_{\mathcal{C}}) + a_s(\mu_0) S_{\gamma g}^{L}(b,\zeta_2,\mu_0).
\end{align}
The functions  $\mathcal{C}$ and $S$ in eq.~(\ref{eq:hats1}-\ref{eq:hats3}) are the result of the $\phi_b$ integration in collinear-soft and dijet soft functions.
For the heavy meson case we have just contributions from gluon scattering, 
\begin{align}\label{eq:siH}
\frac{d\sigma}{d\Pi dr_T}&=\frac{d\sigma^U(\gamma^* g)}{d\Pi dr_T}+\frac{d\sigma^L(\gamma^* g)}{d\Pi dr_T},
\end{align}
and we have to change $J_{f,\bar f}\rightarrow H_+$ and $ \mathcal{C}_f\rightarrow  \mathcal{J}_{Q\rightarrow H} $
in eq.~(\ref{eq:sifg1}-\ref{eq:sifg1L}).
In the case of angular modulation the cross-sections can also be written as in eq.~(\ref{eq:siJ}-\ref{eq:siH}), with the correct  values  of the functions $I_\mathrm{const.}$, $\mathcal{C}$ and $S$.
The non-perturbative effects are in all cases encoded in the evolution kernels, TMD and jet functions.
In the next section we describe  how the evolution kernels are defined.
\section{Evolution kernels and scale choices}
\label{sec:Evolution}
The evolution kernels appearing in eq.~(\ref{eq:siJ}-\ref{eq:siH}) are
\begin{align} \nonumber 
    \mathcal{R}_g \lp (\{\mu_k\},\zeta_{1,0},\zeta_{2,0}) \to (p_T,p_T^2,1) \rp &= \mathcal{R}_{J_f}(\mu_J \to p_T)^2 \mathcal{R}_{\mathcal{C}_f}(\mu_\mathcal{C} \to p_T)^2 \\ & \label{eq:r1}
    \times \mathcal{R}_F^g \lp (\mu_0,\zeta_{1,0}) \to (p_T,p_T^2) \rp \mathcal{R}_S^q \lp (\mu_0,\zeta_{2,0}) \to (p_T,1) \rp, \\ \nonumber
\mathcal{R}_q \lp (\{\mu_k\},\zeta_{1,0},\zeta_{2,0}) \to (p_T,p_T^2,1) \rp &= \mathcal{R}_{J_f}(\mu_J \to p_T)
\mathcal{R}_{J_g}(\mu_J \to p_T)\mathcal{R}_{\mathcal{C}_f}(\mu_\mathcal{C} \to p_T) \mathcal{R}_{\mathcal{C}_g}(\mu_\mathcal{C} \to p_T)\\ &
    \times \mathcal{R}_F^q \lp (\mu_0,\zeta_{1,0}) \to (p_T,p_T^2) \rp \mathcal{R}_S^g \lp (\mu_0,\zeta_{2,0}) \to (p_T,1) \rp, \label{eq:r2}
\end{align}
where  $\mathcal{R}_{J_{f,g}}$ is a jet function kernel, $\mathcal{R}_{\mathcal{C}_{f,g}}$ is the one of collinear-soft functions, $\mathcal{R}_F^{q,g}$ the one of TMD and finally $\mathcal{R}_S^{q g}$ is the one of the dijet soft function. In the heavy quark case the evolution kernels are parameterized like in eq.~(\ref{eq:r1}) with the  usual changes $J_{f}\rightarrow H_+$ and $ \mathcal{C}_f\rightarrow  \mathcal{J}_{Q\rightarrow H} $.
The kernels for single-scale evolution have a standard form and a review up to NLL is given in \cite{Hornig:2016js},
\begin{align}
    \mathcal{R}_{i}(\mu_i \to p_T)&=e^{K_i(\mu_i \to p_T)} \lp \frac{\mu_i}{m_i} \rp^{\omega_i(\mu_i \to p_T)},\qquad i = \{ \mathcal{C}_f, \mathcal{C}_g, J_f, J_g, \mathcal{J}_{Q \to H}, H_{+} \}
\end{align}
where
\begin{align}
\label{eq:omegaFNLL}
 \omega_i(\mu_i \to p_T) \Big\vert_{\rm NLL}  &=-\frac{\Gamma_{i}^0}{\beta_0} \left[\ln{r}+\left(\frac{\Gamma_{1}}{\Gamma_{0}}-\frac{\beta_1}{\beta_0}\right)\frac{\alpha_s(\mu_i)}{4\pi}(r-1)\right] \,,\\
  \label{KFNLL}
K_i(\mu_i \to p_T) \Big\vert_{\rm NLL} &=-\frac{\gamma_{i}^0}{2\beta_0}\ln {r} - \frac{2\pi\Gamma_{i}^0}{(\beta_0)^2}\bigg[\frac{r-1-r\ln{r}}{\alpha_s(p_T)} \nonumber \\
  & \qquad \qquad   +\left(\frac{\Gamma_{1}}{\Gamma_{0}}-\frac{\beta_1}{\beta_0}\right)\frac{1-r+\ln{r}}{4\pi}+\frac{\beta_1}{8\pi\beta_0}\ln^2{r}\bigg],
\end{align}
with $r=\alpha_s(p_T)/\alpha_s(\mu_i)$ and
\begin{align}
    \Gamma^0_{\mathcal{C}_f} & = - 4 C_F,\quad \Gamma^0_{\mathcal{C}_g} = - 4 C_A,\quad \gamma^0_{\mathcal{C}_{f/g}} = 0,\quad m_{\mathcal{C}_{f/g}} = \frac{R e^{-\gamma_E}}{b}, \nonumber \\
    \Gamma^0_{J_f} & = 4 C_F,\quad \Gamma^0_{J_g} = 4 C_A,\quad \gamma^0_{J_{f}} = 6 C_F,\quad \gamma^0_{J_{g}} = 2 \beta_0,\quad m_{J_{f/g}} = p_T R, \nonumber \\
    \Gamma^0_{\mathcal{J}} & = - 4 C_F,\quad \gamma^0_{\mathcal{J}} = 4 C_F,\quad m_{\mathcal{J}} = \frac{m_Q/p_T e^{-\gamma_E}}{b}, \nonumber \\
    \Gamma^0_{+} & = 4 C_F,\quad \gamma^0_{+} = 2 C_F,\quad m_{+} = m_Q,
\end{align}
Initial scales $\mu_i$ choice is given in sec.~\ref{sec:results}. The TMD kernel is considered here in the $\zeta$-prescription described in \cite{Scimemi:2018xaf} and implemented in the code Artemide \cite{web} that we use,
\begin{equation}
    \mathcal{R}_F^{q,g}(\{\mu_0,\zeta_0\}\rightarrow\{\mu_f,\zeta_f\}) = \lp \frac{\zeta_f}{\zeta_{\mu} (b, \mu_f)} \rp^{-\mathcal{D}_{q,g}(b, \mu_f)}.
\end{equation}
 In the next paragraph we define a $\zeta$-prescription also for the dijet evolution kernel $\mathcal{R}_S^g \lp (\mu_0,\zeta_{2,0}) \to (p_T,1)$, which is the only missing part.

\subsection{$\zeta$-prescription for dijet evolution kernel}
The angular independent  kernel of the dijet soft function is obtained as a solution of  a coupled system of differential equations, reported in eq.~(\ref{eq:kernelS}), that are formally very similar to the TMD ones \cite{Collins:2011zzd,GarciaEchevarria:2011rb}. 
 The anomalous dimensions are given by 
\begin{align}
    \bar{\gamma}_{S_{\gamma g}}(\mu,\zeta) & = \gamma_{\mathrm{cusp}} \lb 2 C_F \ln \lp \frac{\mu^2}{\mu_0^2} \rp - C_A \ln \lp \frac{\zeta}{\zeta^{\gamma g}_{2,0}} \rp \rb + \delta \gamma_S^{\gamma g}, \\
    \bar{\gamma}_{S_{\gamma f}}(\mu,\zeta) & = \gamma_{\mathrm{cusp}} \lb (C_F + C_A) \ln \lp \frac{\mu^2}{\mu_0^2} \rp - C_F \ln \lp \frac{\zeta}{\zeta^{\gamma f}_{2,0}} \rp \rb + \delta \gamma_S^{\gamma f},
\end{align}
where
\begin{equation}
    \mu_0 = \frac{2}{b e^{\gamma_E}},\qquad \zeta_{2,0}^{\gamma g} = \lp\frac{4 p_T^2 }{\hat{s}}\rp^{\frac{2C_F}{C_A}}, \qquad \zeta_{2,0}^{\gamma f} = \lp \frac{4 p_T^2}{\hat{s}} \rp^{\frac{C_F+C_A}{C_F}} \lp \frac{\hat{t}}{\hat{u}} \rp^{\frac{C_F-C_A}{C_F}},
\end{equation}
and $\delta \gamma_{S}$ are the non-cusp SF anomalous dimension, which is known up to three-loops for the gluon-channel and up to one-loop for the quark-channel and are reported in appendix.  The anomalous dimension  and the rapidity anomalous dimension (RAD) in eq.~(\ref{eq:AD_evol}, \ref{eq:RAD_evol}) satisfy also  
\begin{align}
  -  \frac{d}{d \ln \zeta} \bar \gamma_{S_{\gamma i}} (\mu, \zeta) =
    \frac{d}{d \ln \mu} \mathcal{D}_{ i} (\mu,b) =  \Gamma_{\mathrm{cusp}}(\mu)  \label{eq:AD-RAD}
\end{align}
The evolution for the SF takes the general form
\begin{equation}
    \mathcal{R}_S^i(\{\mu_i,\zeta_i\}\rightarrow\{\mu_f,\zeta_f\}) = \exp \left[\int_{P} \lp 
    \bar\gamma_{S_\gamma i}(\mu, \zeta) d \ln \mu -\mathcal{D}_i(\mu, b) d \ln \zeta \rp \right]
\end{equation}
with $i=q,g$ and
$\{\mu_i,\zeta_i\}$ and $\{\mu_f,\zeta_f\}$ being the initial and final points of factorization and rapidity scales. 
The integration path $P$ is an arbitrary path in the $\{\mu,\zeta\}$-plane.  Eq.~(\ref{eq:AD-RAD}) ensures that the evolution kernel is path only independent when one knows the complete perturbative expansion of the anomalous dimensions. Since this is not the case
 the path independence is broken. In order to partially restore the path independence we proceed as in \cite{Scimemi:2018xaf} defining a $\zeta$-prescription also for the dijet  soft function evolution kernel.
The $\zeta$-prescription provides a way to choose the initial scale $\zeta_i$ of the evolution kernel as a function of $\mu$ and $b$ so that the SF does not depend on the initial scale $\mu_i$. This is done by taking the integration path through a null-evolution line in the $\{\mu,\zeta\}$-plane and then taking a fixed-$\mu$ evolution.

To find the null-evolution line we interpret the pair of differential equations \eqref{eq:AD-RAD}  as a two-dimensional gradient equation $\bmat \nabla F = \bmat E F$, where $\bmat E = (\gamma_S(\mu,\zeta), -\mathcal{D}_S(\mu, b))$. The null-evolution line is then an equipotential line of the field $\bmat E$. In particular, there is a special null-evolution line that passes through the saddle-point $\{\mu_{\mathrm{saddle}},\zeta_{\mathrm{saddle}}\}$ of the evolution field. We find that the saddle point is exactly $\mu_{\mathrm{saddle}}=\mu_0$ and $\zeta_{\mathrm{saddle}}^{\gamma i}=\zeta_0^{\gamma i}$. If we parameterize the null-evolution line as $\{ \mu,\zeta_{\mu}(b) \}$, the value of $\zeta_{\mu}$ is given by
\begin{equation}\label{eq:zeta_mu_diff_eq}
    \gamma_{S_{\gamma i}}(\mu, \zeta_{\mu}(b)) = 2 \mathcal{D}_{S_{\gamma i}} (\mu,b) \frac{d \ln \zeta_{\mu}(b)}{d\ln \mu^2},
\end{equation}
which is solved perturbatively order by order in $\alpha_s$. The perturbative solution takes the form
\begin{align}
    \zeta_{\mu,\, \mathrm{pert}}^{\gamma g}(b) & = \lp \frac{\mu}{\mu_0} \rp^{\frac{2C_F}{C_A}}\zeta_0^{\gamma g} e^{v^{\gamma g}(\mu,b)},\\
    \zeta_{\mu,\, \mathrm{pert}}^{\gamma f}(b) & = \lp \frac{\mu}{\mu_0} \rp^{\frac{C_F+C_A}{C_F}}\zeta_0^{\gamma f} e^{v^{\gamma f}(\mu,b)},
\end{align}
where
\begin{equation}
    v^{\gamma i}(\mu,b)=\sum_{n=0}^{\infty}a_s^n(\mu) v_n^{\gamma i} (\mathbf{L}_{\mu}),\qquad \mathbf{L}_{\mu} = \ln(B\mu^2 e^{2\gamma_E}), 
\end{equation}
\begin{align}
    v_{0}^{\gamma g}\left(\mathbf{L}_{\mu}\right) & = 0, \\
    v_{1}^{\gamma g}\left(\mathbf{L}_{\mu}\right) & = \frac{2 C_F}{C_A} \lb -\frac{\beta_{0}}{12} \mathbf{L}_{\mu}^{2} + \frac{\frac{\gamma_2}{2C_F} - d^{(2,0)}}{\Gamma_{0}} \rb, \\
    v_{2}^{\gamma g}\left(\mathbf{L}_{\mu}\right) & = \frac{2 C_F}{C_A} \lb -\frac{\beta_{0}^{2}}{24} \mathbf{L}_{\mu}^{3} - \left(\frac{\beta_{1}}{12}+\frac{\beta_{0} \Gamma_{1}}{12\Gamma_{0}}\right) \mathbf{L}_{\mu}^{2}+\left(\frac{\beta_{0} \frac{\gamma_2}{2C_F}}{2 \Gamma_{0}} - \frac{4 \beta_{0} d^{(2,0)}}{3 \Gamma_{0}}\right) \mathbf{L}_{\mu} \nl
    & -\frac{\frac{\gamma_2}{2C_F} \Gamma_{1} - d^{(2,0)} \Gamma_{1}}{\Gamma_{0}^{2}}+\frac{\frac{\gamma_3}{2C_F} - d^{(3,0)}}{\Gamma_{0}} \rb,\\
    v_{0}^{\gamma f}\left(\mathbf{L}_{\mu}\right) & = 0,
\end{align}
and we are using the following notation
\begin{equation}
    \mathcal{D}_i(\mu,b) = C_i\sum_{n=1}^{\infty}a_s^n(\mu)\sum_{k=0}^{n} \mathbf{L}_{\mu}^k d^{(n,k)},\quad \delta\gamma_S(\mu) = \sum_{n=1}^{\infty} a_s^n(\mu) \gamma_n,
\end{equation}
\begin{equation}
    \beta(a_s) = -\sum_{n=0}^{\infty}a_s^{n+2}\beta_n,\quad \Gamma_{\mathrm{cusp}}(\mu) = C_i\gamma_{\mathrm{cusp}}(\mu)  = C_i\sum_{n=0}^{\infty} a_s^{n+1}(\mu)\Gamma_n,
\end{equation}
with $C_i = C_F, C_A$ for quark and gluon channel respectively. Notice that $v_0$ vanishes as it is proportional to the LO non-cusp AD, which is zero for the SF. The non-cusp AD is not known beyond LO for the quark-channel.

The RAD is a function of $b$ and therefore has important non-perturbative corrections in the large-$b$ region. These corrections can be implemented as a model. The way to proceed is to solve \eqref{eq:zeta_mu_diff_eq} for a generic non-perturbative RAD. The equation is solvable but it is difficult to obtain the cancellation of perturbative logarithms in the small-$b$ region. Following \cite{Scimemi:2018xaf} we use the perturbative solution for the small-$b$ region and move to the exact (generic RAD) solution for large-$b$:
\begin{equation}
    \zeta_{\mu}(b)=\zeta_{\mu}^{\mathrm{pert}}(b) e^{-b^{2}/B_{\mathrm{NP}}^2}+\zeta_{\mu}^{\mathrm{exact}}(b)\left(1-e^{-b^{2}/B_{\mathrm{NP}}^{2}}\right),
\end{equation}
with $B_{\mathrm{NP}}$ being the $b$ value where non-perturbative (NP) effects become important ($\sim 2.5$ GeV$^{-1}$). We have already discussed the perturbative solution to eq. \eqref{eq:zeta_mu_diff_eq}. For the exact solution we find
\begin{align}
    \zeta_{\mu,\,\mathrm{exact}}^{\gamma g}(b) & = \lp \frac{\mu^2}{\mu_0^2} \rp^{\frac{2C_F}{C_A}}\zeta_0^{\gamma g} e^{-g^{\gamma g}(a_s,\mathcal{D}_S)/\mathcal{D}_S},\\
    \zeta_{\mu,\,\mathrm{exact}}^{\gamma f}(b) & = \lp \frac{\mu^2}{\mu_0^2} \rp^{\frac{C_F+C_A}{C_F}}\zeta_0^{\gamma f} e^{-g^{\gamma f}(a_s,\mathcal{D}_S)/\mathcal{D}_S},
\end{align}
where
\begin{equation}
    g^{\gamma i}(a_s, \mathcal{D}_S) = \frac{1}{a_{s}} \frac{\Gamma_{0}}{2 \beta_{0}^{2}} \sum_{n=0}^{\infty} a_{s}^{n} g_{n}^{\gamma i}(\mathcal{D}_S),
\end{equation}
\begin{align}
    g^{\gamma g}_{0} & = \frac{2 C_F}{C_A}\lb e^{-p}-1+p \rb,\\
    g_{1}^{\gamma g} & = \frac{2 C_F}{C_A} \lb \frac{\beta_{1}}{\beta_{0}}\left(e^{-p}-1+p-\frac{p^{2}}{2}\right)-\frac{\Gamma_{1}}{\Gamma_{0}}\left(e^{-p}-1+p\right) \rb, \\
    g^{\gamma f}_{0} & = \frac{C_F + C_A}{C_F}\lb e^{-p}-1+p \rb,\\
    g_{1}^{\gamma f} & = \frac{C_F + C_A}{C_F} \lb \frac{\beta_{1}}{\beta_{0}}\left(e^{-p}-1+p-\frac{p^{2}}{2}\right)-\frac{\Gamma_{1}}{\Gamma_{0}}\left(e^{-p}-1+p\right) \rb,
\end{align}
and $p=2\beta_0\mathcal{D}_S/\Gamma_0$.

Finally, the evolution kernel that provides the evolution from the null-evolution line and that passes through the saddle-point to the final $\zeta$ point is given by
\begin{equation}
    \mathcal{R}_S^{q,g}(\{\mu_0,\zeta_0\}\rightarrow\{\mu_f,\zeta_f\}) = \lp \frac{\zeta_f}{\zeta_{\mu} (b, \mu_f)} \rp^{-\mathcal{D}_{q,g}(b, \mu_f)},
\end{equation}
and if we consider the evolution from an arbitrary initial scale we take
\begin{equation}
    \mathcal{R}_S^i(\{\mu_i,\zeta_i\}\rightarrow\{\mu_f,\zeta_f\}) = \frac{\mathcal{R}_S^i(\{\mu_0,\zeta_0\}\rightarrow\{\mu_f,\zeta_f\})}{\mathcal{R}_S^i(\{\mu_0,\zeta_0\}\rightarrow\{\mu_i,\zeta_i\})}.
\end{equation}
with $i=q,g$.
This discussion concludes the analysis of all terms that appear in the factorization theorem and the scale prescription. We are now ready for the implementation in the code Artemide \cite{web}.



\section{Dijet and heavy hadron pair (HHP) production at EIC}
\label{sec:results}

In order to test the phenomenology developed in the previous sections we consider the case of the EIC.
In \cite{delCastillo:2020omr} we already studied the coverage of the EIC and we concluded that the most favourable case is given for a value of mass energy for dijet production around $\sqrt{s} = 140\, \text{GeV}$ and central rapidity, $\eta_1 = \eta_2 = 0$. Typical values for jet radii and momenta at EIC are  respectively $R\sim 0.7$ and $p_T\sim Q/2 \sim 20$ GeV.
In order to simplify the discussion we show plots integrated over Bjorken variable $x$ (the longitudinal fraction of momentum $\xi$ that enters in the TMDPDFs is  $\xi \sim 2x$) in the allowed kinematic intervals. 
For the case of central rapidity we have $x \in (0.0859,0.5)$.
The cross-sections that we plot are
\begin{align}
    \int_{x_{\rm min}}^{x_{\rm max}} dx   \frac{d\sigma}{dx d\eta_1 d\eta_2 d p_T d r_T}\Big|_{\eta_1 ,\; \eta_2,\; p_T}
\end{align}
and its value is presented as a function of the small transverse momentum $r_T$.
The cross-sections and the error bands are obtained by using and preparing specific moduli for the code Artemide \cite{web}.  In particular we use the TMD and the TMD evolution kernels already coded in Artemide, that come from the fit \cite{Scimemi:2019cmh},  while the new functions studied in this work are included in this code for the first time. 
The gluon TMD is not fitted yet, however in Artemide there is a parameterization for it.
The code takes into account that the contribution of linearly polarized gluons is highly suppressed because in the the small-$b$ regime the matching of the linearly polarized gluon TMD onto the gluon PDF starts at order $\alpha_s^1$ and not at order $\alpha_s^0$ like other distributions. 
In ref.~\cite{Gutierrez-Reyes:2019rug} the cross section obtained in this way agrees with Pythya 8 and current experimental results for the Higgs transverse momentum spectrum, which are however not very precise. 
The non-perturbative effects are expected to be important in the high-$b$ region and they should not alter the small-$b$ behavior of this distribution. 
Notice also that the non-perturbative effects play a role to control the behavior of the distribution around the Landau pole at large-$b$, which means a further suppression effect at large-$b$ (as we also observe in the case of unpolarized distributions). Summing up, given the current perturbative and non-perturbative knowledge of TMDs, at this stage we prefer not to push for a hypothetical non-perturbative enhancement of the contribution of linearly polarized gluon TMD.

\begin{figure}
\begin{center}
\includegraphics[width=0.8\textwidth]{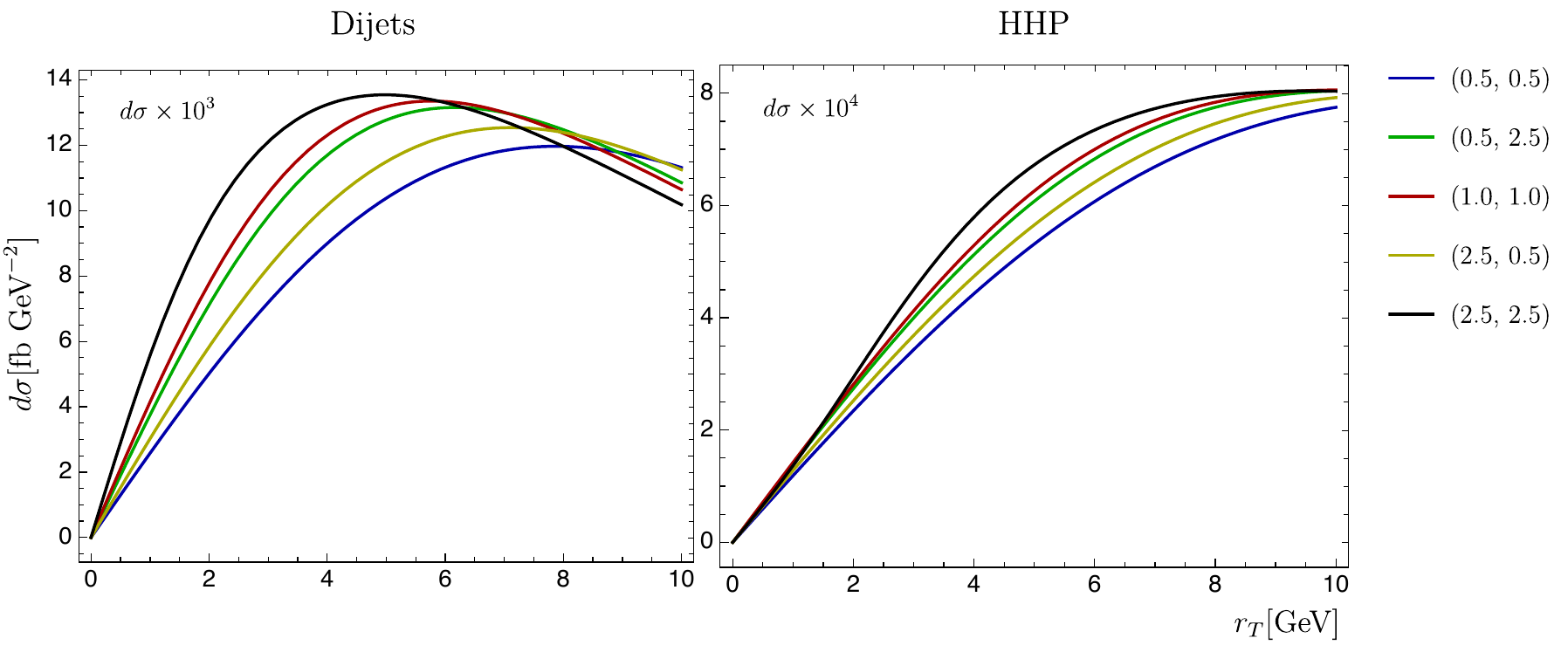}
\caption{Impact of $B_{\mathrm{NP}}$ variation over dijets and heavy meson total cross-section. Legend correspond to $(B_{\mathrm{NP}}^{S},B_{\mathrm{NP}}^{\mathcal{C}})$ and $(B_{\mathrm{NP}}^{S},B_{\mathrm{NP}}^{\mathcal{J}})$ for dijet and HHP production respectively}
\label{fig:BNP}    
\end{center}
\end{figure}

\begin{table}[h!]
    \centering
    \begin{tabular}{|c||c|c|c|}
    \hline
            & $\mathcal{C}$ & $\mathcal{J}$ & $S$  \\
         \hline
     $B_{\mathrm{NP}}^i$ (GeV$^{-1}$) & 2.5 & 2.5 & 2.5  \\
    \hline
    \end{tabular}
     \begin{tabular}{|c||c|c|}
    \hline
           & $\mathcal{C}$ & $\mathcal{J}$   \\
         \hline
    $b_{\mathrm{max}}$ (GeV$^{-1}$)  & 0.5 & 0.3   \\
    \hline
    \end{tabular}
    \caption{\label{tab1}Values of non-perturbative parameter $B_{\mathrm{NP}}$ and $b_{\mathrm{max}}$ prescription chosen for collinear-soft function, heavy meson jet function and dijet soft function. Impact of the varition of $B_{\mathrm{NP}}$ is shown in fig.~\ref{fig:BNP}.}
\end{table}
The factorization that we propose in general needs information of the non-perturbative effects in several functions.
For the dijet case we have
\begin{align}
    \mathcal{C}(b, R; p_T) & = \mathcal{R}_{\mathcal{C}}(b,R; p_T, \mu_{\mathcal{C}})\mathcal{C}^{\it pert}
    (b, R; \mu_{\mathcal{C}}) f^{\mathrm{NP}}_{\mathcal{C}}(b, R), \\
     S_{\gamma i}(b; p_T,1) & = \mathcal{R}_S(\{\mu_0,\zeta_0\}\rightarrow\{p_T,1\})S_{\gamma i}^{\it pert}(b; \mu_0, \zeta_0)
      f^{\mathrm{NP}}_{\mathcal{S}}(b), 
\end{align}
where
the functions with suffix {\it pert} refer to their perturbative part in the $\overline{\rm MS}$ scheme  which is currently known at one loop.
 Similarly for the HHP case we need
\begin{align}
\mathcal{J}(b, m_Q/p_T; p_T) & = \mathcal{R}_{\mathcal{J}}(b,m_Q/p_T; p_T, \mu_{\mathcal{J}})\mathcal{J}^{\it pert}
    (b, m_Q/p_T; \mu_{\mathcal{J}}) f^{\mathrm{NP}}_{\mathcal{J}}(b;m_Q).
    \end{align}
 The non-perturbative effects are parameterized as 
\begin{equation}
    f^{\mathrm{NP}}_{i}(b)=\exp \lp -\frac{b^2}{(B_{\mathrm{NP}}^i)^2} \rp,\quad\quad i=\mathcal{C},\mathcal{J}, S.
\end{equation}
The values of $B_{\mathrm{NP}}^i$ define thee non-perturbative model and we have tested several combinations as  shown in fig.~\ref{fig:BNP}.
Higher values of $B_{\mathrm{NP}}^i$ are more sensitive to the perturbative series in the low transverse momentum spectrum,  and in general provide higher values of the observables. In unpolarized TMD cases we have usually that typical values of $B_{\mathrm{NP}}^i$ are around 1-3 GeV$^{-1}$ so we have found reasonable to fix their values as in tab.~\ref{tab1}.

The  factorization scales $\mu_{\mathcal{C}}$ for dijet and $\mu_{\mathcal{J}}$ for HHP are chosen to minimize  perturbative logarithms and to not hit the Landau pole of the strong coupling constant, 
\begin{align}
    \mu_{\mathcal{C}} = 2 e^{-\gamma_E} & \lp \frac{1}{b} + \frac{1}{b_{\mathrm{max}}} \rp, \label{eq:muC}\\
    \mu_{\mathcal{J}} = \frac{1}{2}e^{-\gamma_E} & \lp \frac{1}{b} + \frac{1}{b_{\mathrm{max}}} \rp.\label{eq:muJ}
\end{align}
This scale choice deserves some comments.
In the dijet case the scale choice does not
 include the dependence on the jet radius $R$. Similarly, the mass of the ratio $m_Q/p_T$ does not enter the collinear-soft function and heavy meson jet. In all cases, this means that there is not a complete cancellation of the logarithms of these functions. The reason is that the $\phi_b$ integration imposes some constraints on the choice of scales.  In fact, the function $\mathcal{A}(\{\mu_i\})$ defined in eq.~(\ref{eq:A_def}), which depends on the initial scale choice for the soft function, collinear-soft function and heavy meson jet, needs to be $\mathcal{A}>-1/2$ in order to have a well defined angular integration. Because of this constraint some scale choices which could be considered like for instance 
\begin{align}
    \mu_{\mathcal{C}} = \frac{R e^{-\gamma_E}}{b},\qquad
    \mu_{\mathcal{J}} = \frac{m_Q/p_T e^{-\gamma_E}}{b},
\end{align}
can not be used. As a result in
our approach we only partially resum the logs in the collinear-soft function and the heavy meson jet in order to maintain the structure of $\zeta$-prescription and double scale evolution in the soft function that is described in sec.~\ref{sec:Evolution}. This leads to the initial scales in eq.~(\ref{eq:muC}, \ref{eq:muJ}).

Finally for the dijet soft function we use the $b^*$-prescription in the same way as for the TMDPDF:
\begin{equation}
    \mu_{S} = \frac{2 e^{-\gamma_E}}{b^*},\quad b^* = \frac{b}{\sqrt{1+b^2/b_{\mathrm{max}}^2}}.
\end{equation}

Concerning the theoretical errors, the scale variations in collinear-soft and heavy meson jet function are the main source. This is due to the non-cancellation of logs in the functions by the choice of the initial scales. The choice of the values $b_{\mathrm{max}}$ for collinear-soft function and heavy meson jet account for the convergence of our perturbative result. A more consistent way to treat the resummation of these scales is left for a future work, involving the refactorization of these functions.

 For functions that do not depend on $b$  the initial scale choice does not require a prescription or NP-model and it is dictated by the cancellation of  the logarithms. For the jet function and the $H_+$ matching coefficient we have
\begin{equation}
    \mu_J = p_T R,\qquad \mu_+ = m_Q.
\end{equation}

We use a NP-model for the rapidity anomalous dimension that enters the exact solution for the null-evolution $\zeta_\mu$ line as it is explained in sec.~\ref{sec:Evolution}. In particular, we use the same model that has been used for TMDPDF in \cite{Scimemi:2019cmh}
\begin{equation}
    \mathcal{D}_{F,S}^{\mathrm{NP}} = c_0 b b^*, \qquad c_0 = 2.5\cdot 10^{-2}.
\end{equation}
This model dictates how the rapidity anomalous dimension behaves in the large-$b$ region and is used for both dijet soft function and TMDPDF when performing double scale evolution. While a color re-scaling of the non-perturbative models for gluon TMDPDF and gluon channel soft function with respect to their quark analogues is possible, we observe that this change does not have a significant impact on the cross-section and, therefore, we choose to keep the same model for both quarks and gluons.
\subsection{Results}
In this section we show our results for the differential cross-section for both dijet and heavy hadron pair production processes. 
Differential cross-sections are shown with error bands coming from scale variation of the different final and initial scales of the functions appearing in our factorization formulas. 
Scale variation bands are obtained by changing the considered scale by a factor of 2 up and down relative to its central value.

\subsubsection{Results for dijet production}
\begin{figure}
     \centering
     \begin{subfigure}[b]{0.3\textwidth}
         \centering
         \includegraphics[scale=0.6]{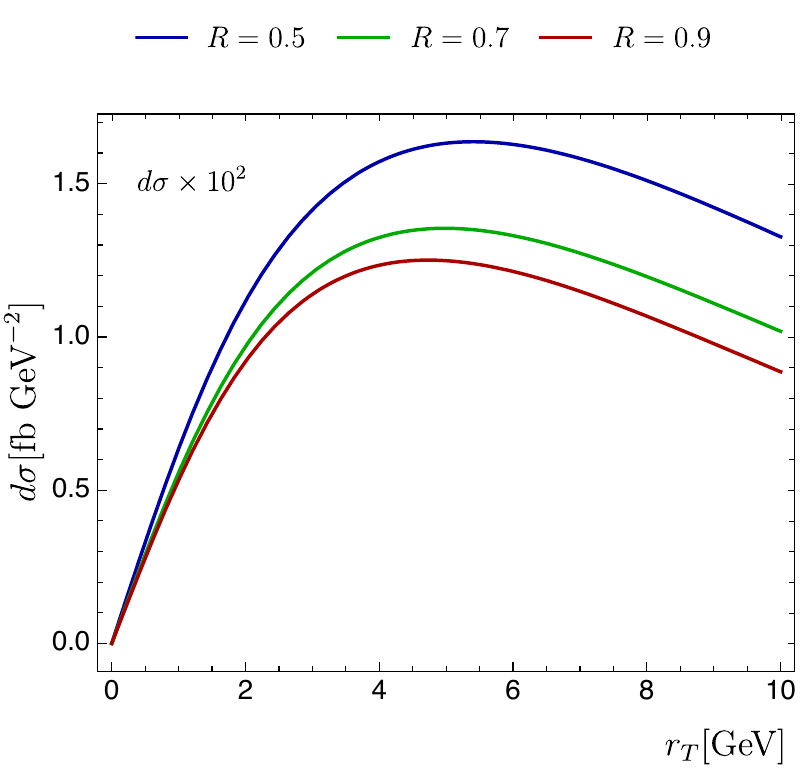}
         \caption{$R$ variation}
         \label{fig:R}
     \end{subfigure}
     \begin{subfigure}[b]{0.3\textwidth}
         \centering
         \includegraphics[scale=0.6]{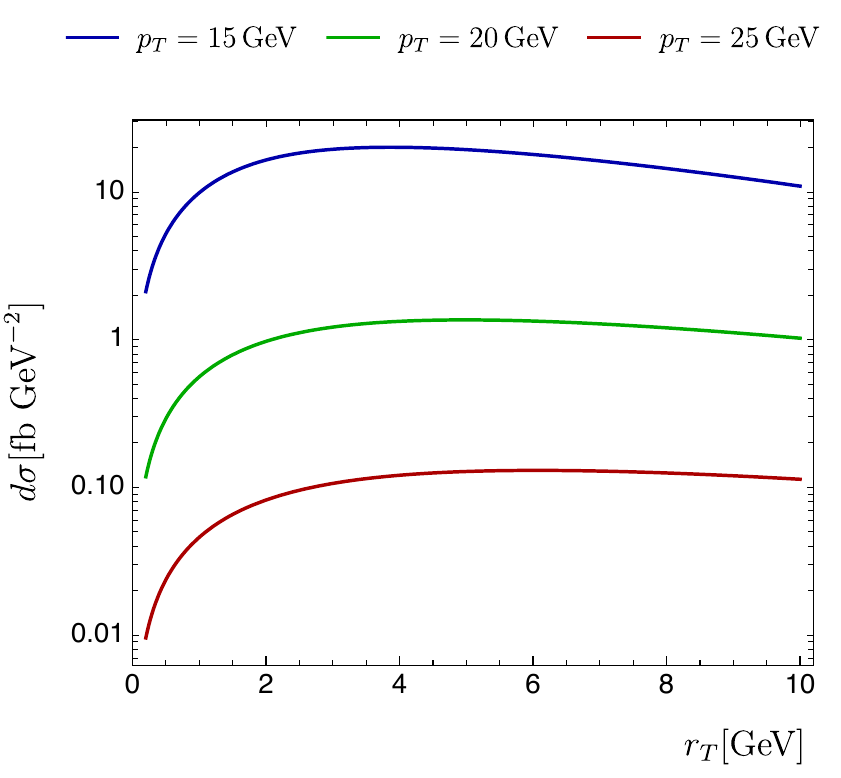}
         \caption{$p_T$ variation}
         \label{fig:pT}
     \end{subfigure}
     \begin{subfigure}[b]{0.3\textwidth}
         \centering
         \includegraphics[scale=0.6]{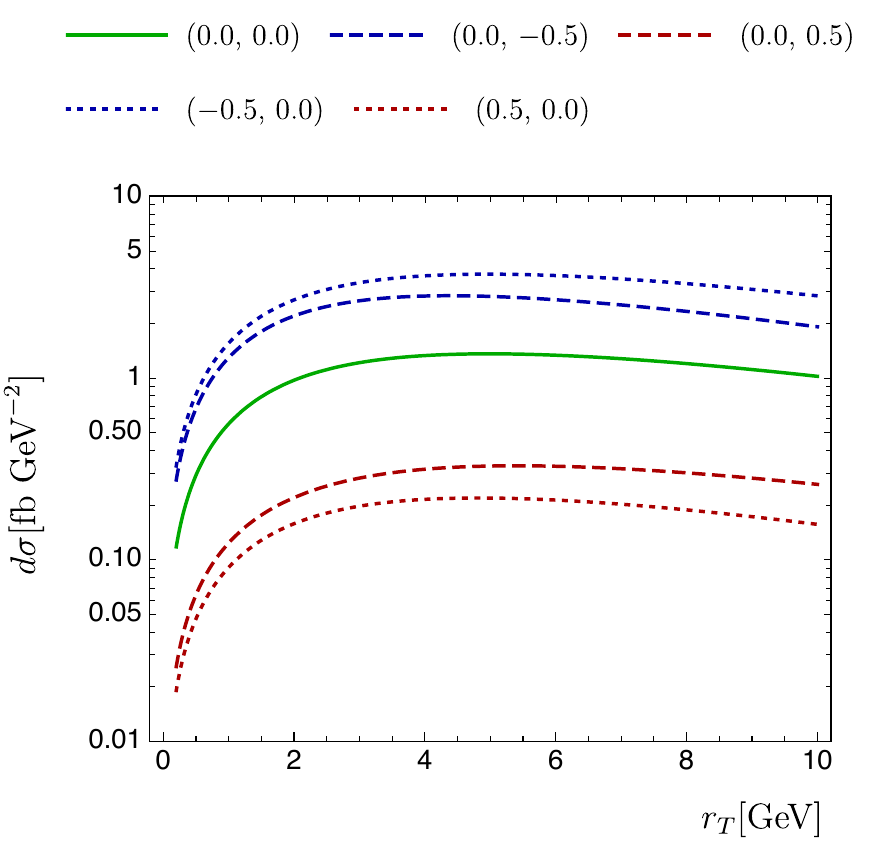}
         \caption{$\eta$ variation $(\eta_1,\eta_2)$}
         \label{fig:eta}
     \end{subfigure}
        \caption{Impact of the variation of the jet radius ($R$), hard scale jet transverse momentum ($p_T$) and jet pseudorapidity ($\eta_i$) for dijet production. For pseudorapidity variation legend is shown referring to $(\eta_1,\eta_2)$ pair, dashed and dotted lines correspond to negative and positive rapidity respectively.}
        \label{fig:variations}
\end{figure}
\begin{figure}
\begin{center}
\includegraphics[width=\textwidth]{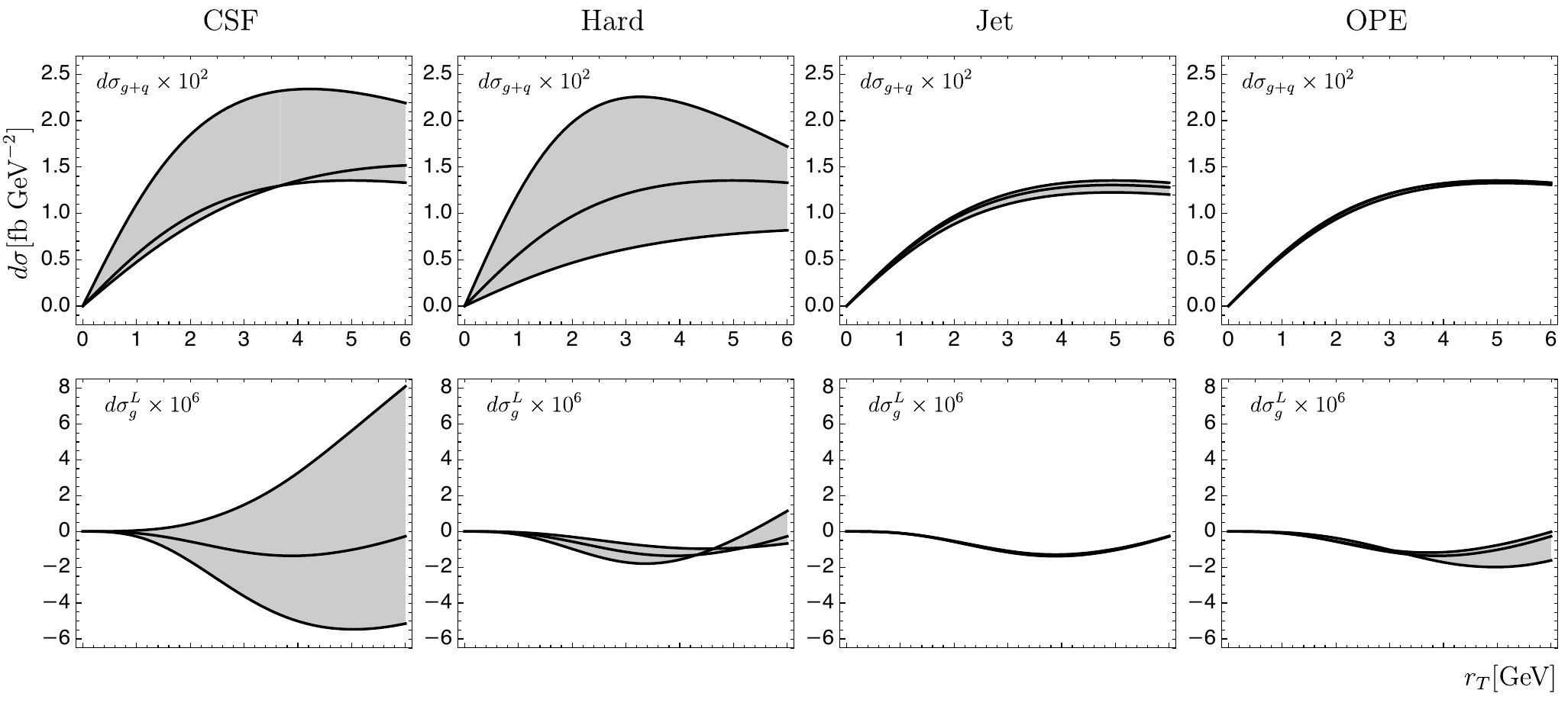}
\caption{Cross-sections for dijet production at EIC with error-bands coming from scale dependence in collinear-soft factor (CSF), hard factor (Hard), jet distributions (Jet) and Wilson coefficients (OPE). Rows correspond to contributions from linearly polarized gluons (top) and total cross-section (bottom). $\sqrt{s} = 140$ GeV, $R = 0.7$, $p_T=20$ GeV, $\eta_1=\eta_2=0$.}
\label{fig:dijets}    
\end{center}
\end{figure}
\begin{figure}
\begin{center}
\includegraphics[width=\textwidth]{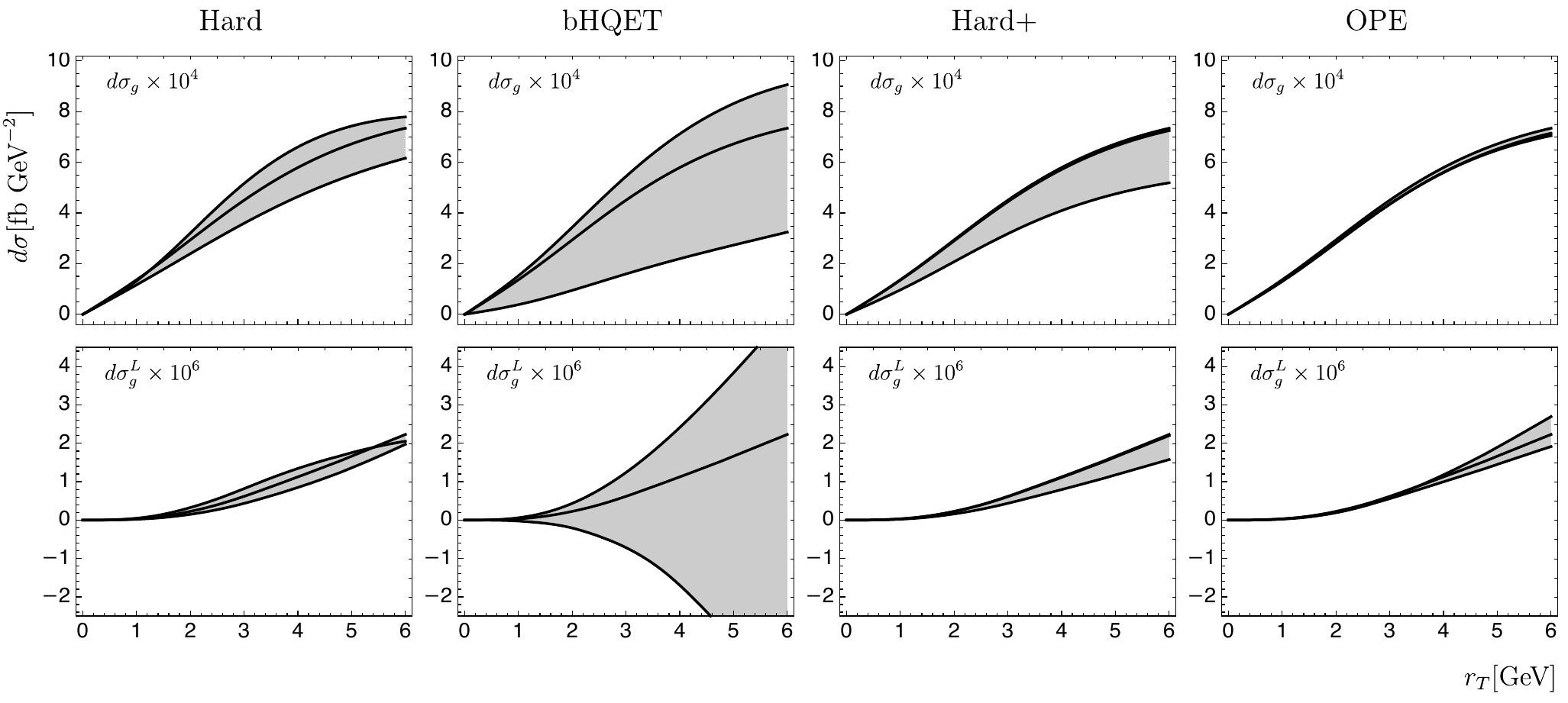}
\caption{Cross-sections for HHP production at EIC with error-bands coming from scale dependence in hard factor (Hard), heavy meson jet function (bHQET), heavy meson jet function matching coefficient (Hard+) and Wilson coefficients (OPE). The rows correspond to contributions from total cross-section (top) and linearly polarized gluons (bottom). $\sqrt{s} = 140$ GeV, $p_T=20$ GeV, $\eta_1=\eta_2=0$.}
\label{fig:heavy}    
\end{center}
\end{figure}
\begin{figure}
\begin{center}
\includegraphics[width=\textwidth]{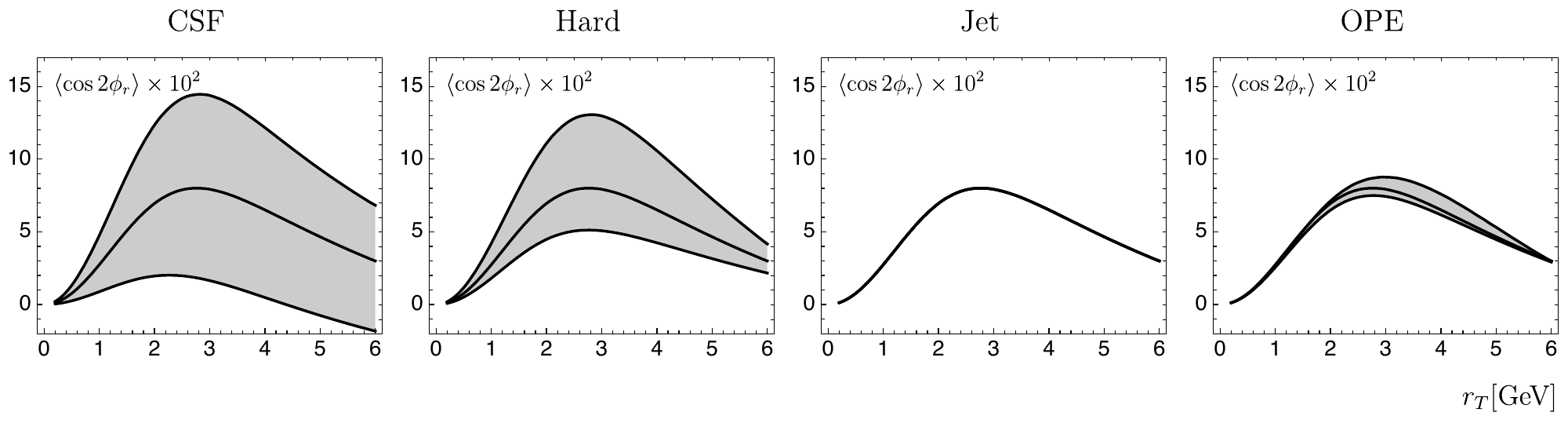}
\caption{Angular modulation contribution for dijet production at EIC with error-bands coming from scale dependence in collinear-soft factor (CSF), hard factor (Hard), jet distributions (Jet) and Wilson coefficients (OPE). $\sqrt{s} = 140$ GeV, $R = 0.7$, $p_T=20$ GeV, $\eta_1=\eta_2=0$.}
\label{fig:AM_dijets}

\end{center}
\end{figure}
\begin{figure}
\begin{center}
\includegraphics[width=\textwidth]{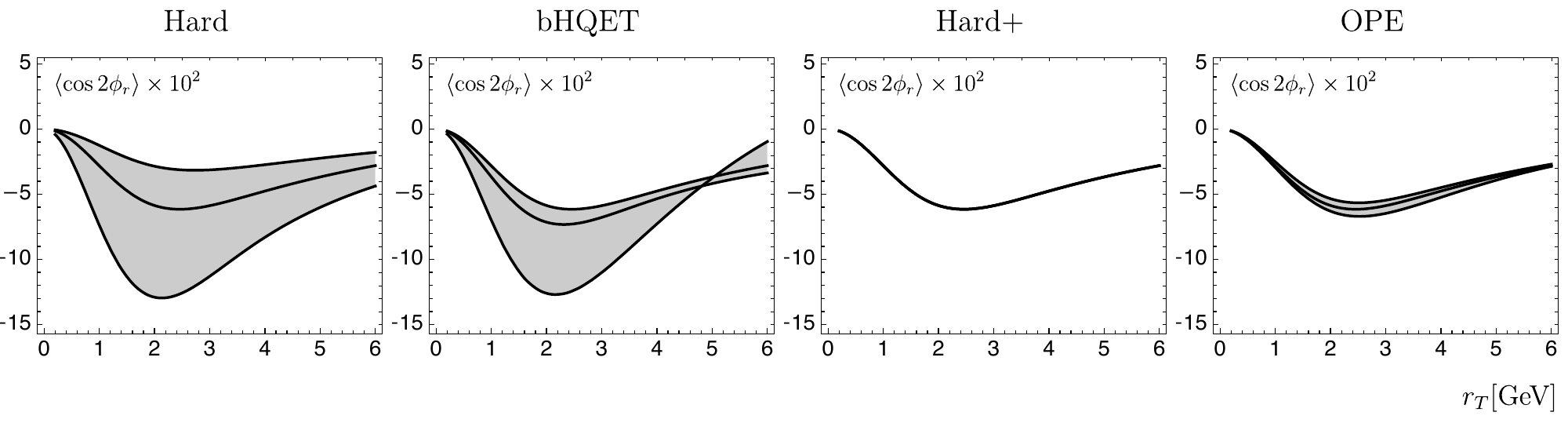}
\caption{Angular modulation contribution for HHP production at EIC with error-bands coming from scale dependence in hard factor (Hard), heavy meson jet function (bHQET), heavy meson jet function matching coefficient (Hard+) and Wilson coefficients (OPE). $\sqrt{s} = 140$ GeV, $p_T=20$ GeV, $\eta_1=\eta_2=0$.}
\label{fig:AM_heavy}    
\end{center}
\end{figure}

In fig.~\ref{fig:variations} we show the impact of the change of jet radius, jet transverse momentum (hard scale) and jet pseudorapidity over total dijet cross-section. We show that for the variation of the jet radius we see a change of around $20\%$ on the cross-section from the central value when taking the jet radius to be $\pm0.2$ from $R=0.7$. For $p_T$ there is a variation of an order of magnitude in the total cross-section when taking $\pm5$ GeV from $20$ GeV. This corresponds to $Q=30,40,50$ GeV respectively. Finally, for pseudorapidity variation we obtain an order of magnitude difference above and below when compared to the central rapidity case. Positive rapidities ($\eta_i=0.5$) correspond to $Q\simeq 53$ GeV while negative rapidities ($\eta_i=-0.5$) correspond to $Q\simeq 32$ GeV, so both $p_T$ and $\eta$ plots are consistent. Notice that total dijet cross-section is not symmetrical for both jet rapidities as for quark channel we have both a quark and gluon jet in the final state. Every other plot is obtained taking $R=0.7$, $p_T = 20$ GeV and $\eta_i=0$.

In fig.~\ref{fig:dijets} the result for the cross-section including quark and gluon channels is shown. We consider the contribution of linearly polarized gluons in a separate panel to show that their contribution is completely negligible,   being a factor $10^{3}$-$10^{4}$ smaller. This leads to the conclusion that the contribution from the linearly polarized gluons can be neglected when considering the unpolarized cross-section.

The angular modulation asymmetry is shown in fig.~\ref{fig:AM_dijets},
being around 5$\%$.

\subsubsection{Results for heavy hadron production}

The analysis for HHP has followed similar steps of the dijet case when possible. The differential cross-section including all channels is plotted in fig.~\ref{fig:heavy}. A separate analysis of the contribution
of linearly polarized gluons show also in this case that they are completely negligible being suppressed by a factor $10^{2}$-$10^{3}$.
The angular modulation asymmetry is shown in fig.~\ref{fig:AM_heavy},
being around 5$\%$.

\section{Conclusions}
Dijet and HHP in SIDIS experiments present an opportunity to study the  gluon TMD. In this work we have considered the case of the Electron Ion Collider (EIC), as an example. The processes have been proven to factorize consistently and this result has been checked at least at one loop \cite{delCastillo:2020omr}. Nevertheless the evolution of the functions that appear in the factorization theorem is non-trivial and we  propose an original solution, which is generic and independent of the resummation framework. We also note that it is consistent with the $\zeta$-prescription of TMD \cite{Scimemi:2018xaf} which we implement in this work. The used prescription allows to separate the evolution kernels from other scale independent factors in the cross-section, so that in our final computations we can use some results already coded in the literature. This is the case for the TMD  and their respective evolution kernels extracted from DY and SIDIS data and presented in the code Artemide \cite{web}.

The phenomenological analysis that we have performed has revealed several issues that need further study in the future.
Estimating the errors due to scale dependence, we have found that several functions need a higher loop calculation to achieve sufficient precision for the low energy jets will be available at EIC. This is the case for instance of the collinear-soft function that appears in the dijet process and more urgently on the heavy hadron jet function. The fact that the perturbative convergence of these function is limited to small values of the $b$ may lead to consider also a re-factorization of these functions, such that the small-$b$ effects are separately resummed. This possibility can eventually be considered in future works.

In all cross-sections we have found a contribution of unpolarized and linearly polarized gluon TMD. For both of these distributions we have used their re-factorization in coefficient functions and collinear PDF studied at higher loops in the literature \cite{Echevarria:2016scs,Gutierrez-Reyes:2019rug,Luo:2019bmw,Luo:2020epw}. In accordance to this well tested procedure, the linearly polarized gluon contribution results to be particularly suppressed in all considered cases, because its matching to collinear PDF starts at order $\alpha_s^1$, instead of $\alpha_s^0$ like the unpolarized distributions. The effect of this suppression is particularly evident in the estimate of the angular modulation of $\cos 2\phi_r$ asymmetry that is here estimated to be around $5\%$. The study of next-to-leading power effects is beyond the purpose of the present work, so that further study is necessary to confirm a value of this asymmetry. 

A source of uncertainty in our prediction comes from the usage of models for many functions that have not been yet compared against data. In this case we have studied several possibilities with simple Gaussian models, assigning values to the non-perturbative parameters according to an educated-guess. The models do not alter the overall-conclusions about scale choices or precision, but can have some effect on the shape of the curves that we have computed. Only an strict comparison with data  or eventual lattice calculations can finally resolve this issue.

As a result of this study we can see that the extraction of gluon TMDs from dijet and HHP processes at the EIC is conditioned yet by the possible theoretical and experimental precision.
In particular, the linearly polarized gluon TMD appears generally too suppressed and hardly accessible if one uses the usual matching of TMDs onto their collinear counterpart distributions.
Nevertheless, the discussed theoretical issues can potentially be solved or improved in future studies.

\section*{Acknowledgements}
 R.F.C., M.G.E. and I.S. are supported by the Spanish Ministry grant PID2019-106080GB-C21. This project has received funding from the European Union Horizon 2020 research and innovation program under grant agreement Num. 824093 (STRONG-2020). 
 M.G.E. is supported by the Community of Madrid and UAH joint grant CM/BG/2021-002 (MultiNuS) within the agreement to fund \emph{Beatriz Galindo} researchers. 
 Y.M. is supported by the European Union’s Horizon 2020 research and innovation program under the Marie Sk\l{}odowska-Curie grant agreement No. 754496-FELLINI.
\appendix
\section{Hard prefactors}
\label{app:Prefactors}
The hard prefactors for each channel are given in ref.~\cite{Boer:2016fqd}. We include in this section the ones relevant for our cases
\begin{align}
    \sigma_0^{g U} = 2 \pi p_T \frac{\mathcal{N}}{x s} \frac{A_0^{gU}}{ f_1^g(\xi, \bmat{r}_T)},\qquad \sigma_0^{f U}=2 \pi p_T \frac{\mathcal{N}}{x s} \frac{A_0^{fU}}{ f_1^f(\xi, \bmat{r}_T)} ,\qquad     \sigma_0^{g L} = -4\pi p_T \frac{\mathcal{N}}{x s} \frac{B_2}{ h_1^{\perp}(\xi, \bmat{r}_T)}, 
\end{align}
where
\begin{equation}
    \mathcal{N}=\frac{\alpha^{2} \alpha_{s}}{\pi s p_T^{2}} \frac{1}{x y^{2}},
\end{equation}
\begin{align}
    A_0^{gU} & = e_q^2 T_R \lb \lp 1 + (1-y^2) \rp A_{U+L}^{gU} - y^2 A_{L}^{gU} \rb f^g_1(\xi, \bmat{r}_T), \\
    A_0^{fU} & = e_q^2 C_F \lb \lp 1 + (1-y^2) \rp A_{U+L}^{fU} - y^2 A_{L}^{fU} \rb f^q_1(\xi, \bmat{r}_T),\\
    B_{2}&= e_{q}^{2} T_{R} \lb \lp 1 + (1-y^2) \rp B_{U+L} - y^2 B_{L} \rb \frac{r_T^2}{M_{p}^{2}} h_{1}^{\perp g}\left(\xi, \bmat r_T \right),
\end{align}
and $A$ and $B$ factors are given by
\begin{align}
    A_{U+L}^{fU}&=\frac{1-z}{D^{2}}\left\{1+z^{2}+\left[2 z(1-z)+4 z^{2}(1-z)^{2}\right] \frac{Q^{2}}{p_T^2}+\left[z^{2}(1-z)^{2}\right]\left[1+(1-z)^{2}\right] \frac{Q^{4}}{p_T^4}\right\},\\
    A_{U+L}^{gU}&=\frac{1}{D^{3}}-\frac{z(1-z)}{D^{3}}\left\{2-8 z(1-z) \frac{Q^{2}}{p_T^{2}}\right.
    \left.-z(1-z)[1-2 z(1-z)] \frac{Q^{4}}{p_T^{4}}\right\},\\
    B_{U+L}&=\frac{z(1-z)}{D^{3}}\left\{[1-6 z(1-z)] \frac{Q^{2}}{p_T^{2}}\right\},\\
    A_{L}^{fU}&=4 \frac{z^{2}(1-z)^{3}}{D^{2}} \frac{Q^{2}}{p_T^{2}},\\
    A_{L}^{gU}&=8 \frac{z^{2}(1-z)^{2}}{D^{3}} \frac{Q^{2}}{p_T^{2}},\\
    B_{L}&=-4 \frac{z^{2}(1-z)^{2}}{D^{3}} \frac{Q^{2}}{p_T^{2}},
\end{align}
where $D$ is defined as
\begin{equation}
D=1+z(1-z) \frac{Q^{2}}{p_T^{2}}.
\end{equation}
\section{Anomalous dimensions}
\label{app:AD}
For the two channels in the dijet process the relevant anomalous dimensions up to one-loop are,
\begin{align} 
\gamma_{H_{\gamma g} }^{[1]} &= 4 \lbc   C_F \lb \ln \lp  \frac{ \hat{s} ^2}{\mu^4} \rp  -2 \gamma_q   \rb  + C_A  \ln \lp \frac{\hat{t} \,\hat{u} } {\hat{s} \mu^2} \rp  \rbc \,, \nl
\gamma_{H_{\gamma f} }^{[1]}  & = 4 \lbc  C_F \lb \ln\lp  \frac{\hat{u}^2}{\mu^4}\rp  - 2 \gamma_q  \rb  + C_A   \ln \lp \frac{\hat{s} \, \hat{t}}{\hat{u}\, \mu^2} \rp   \rbc \,,\nl
\gamma_{S_{\gamma g}} ^{[1]}  &= 4 \lbc  - C_A \ln \zeta_2  +  2 C_F \lb \ln (B \mu^2 \,e^{2\gamma_E}) - \ln \hat{s} +\ln p_T^2 +\ln(4 c_{\bmat{b}}^2 )\rb \rbc\,, \nl
\gamma_{S_{\gamma f}} ^{[1]}   & =  4 \lbc  (C_F +C_A) \lb  \ln (B  \mu^2 e^{2\gamma_E})  -\ln \hat{s} +\ln  p_T^2  + \ln (4 c_{\bmat{b}}^2 ) \rb  + (C_F-C_A)\lb  \ln \lp \frac{\hat{t} }{ \hat{u}  }  \rp   -  \kappa(v_f) \rb- C_F \ln\zeta_2    \rbc   \nl
\gamma_{F_i}^{[1]}       &=4 C_i \lb - \ln\lp \frac{\zeta_1}{\mu^2} \rp + \gamma_i \rb \,,\nl
\gamma_{J_i}^{[1]}   &= 4 C_i \lb   -\ln \lp \frac{p_T^2}{\mu^2} \rp  -\ln R^2 + \gamma_i  \rb\,, \nl
\gamma_{\mathcal{C}_g}^{[1]}&= 4 C_A \lb   -  \ln \lp B \mu^2 \,e^{2\gamma_E} \rp  + \ln R^2 -\ln (4 c_{\bmat{b}}^2 )  + \kappa(v_g)\rb \,,\nl
\gamma_{\mathcal{C}_i}^{[1]}&= 4 C_F \lb   -  \ln \lp B \mu^2 \,e^{2\gamma_E} \rp  + \ln R^2 -\ln (4 c_{\bmat{b}}^2 )  + \kappa(v_i)\rb \,,\nonumber \\[8pt]
\gamma_\alpha^{[1]} &= - 4 C_A \gamma_g\,,
\end{align} 
The imaginary component in the soft and collinear-soft anomalous dimension is denoted by $\kappa(v_i)$ where 
\begin{align}
\kappa(v_f) = - \kappa(v_{\bar{f}}) = - \kappa(v_g) = i \pi\, \text{sign}( c_{\bmat{b}}).
\end{align}
These anomalous dimensions, except the soft function which we calculated here, can be found in \cite{Becher:2009th,Becher:2012xr,Chien:2020hzh,Hornig:2016ahz,Buffing:2018ggv,Echevarria:2015byo}. We also  expand $A_{\bmat{b}}$ in the soft function anomalous dimension in terms of $\hat{s}$, $p_T$, and $c_{\bmat{b}}$. It is now easy to confirm the cancelation of the anomalous dimensions at $\mathcal{O}(\alpha_s)$ which also serves as confirmation of the factorization theorem at the same order. 

For the heavy hadron pair case the
one-loop hard function, $H_+$ is, 
\begin{equation}
\label{eq:h-nlo}
H_+ (m_Q,\mu) = 1+ \frac{\alpha_s }{4\pi}C_F \lbc \ln \lp\frac{\mu^2}{m_Q^2} \rp + \ln^2 \lp\frac{\mu^2}{m_Q^2} \rp + 8 + \frac{\pi^2}{6} \rbc ,
\end{equation}
and the corresponding anomalous dimension is 
\begin{equation}
\gamma_{+} = \frac{\alpha_s C_F}{\pi} \lbc  \frac{1}{2 }- \ln \lp \frac{m_Q^2}{\mu^2} \rp \rbc\;.
\end{equation}
The heavy jet functions anomalous dimension is 
\begin{equation}
\gamma_{\mathcal{J}} = \frac{\alpha_s C_F}{\pi} \lbc  1- 2 \ln \mathcal{R} \rbc
\end{equation}
where
\begin{equation}
 \mathcal{R} = -  \frac{i\,p_T \mu \, e^{\gamma_E} (\bmat{v}\cdot \bmat{b} ) }{m_Q |\bmat{v}| } .
\end{equation}

\bibliographystyle{JHEP}
\normalbaselines 
\bibliography{TMD_ref}

\end{document}